\documentclass[usenatbib]{mn2e}
\usepackage{graphicx}
\usepackage{natbib}
\usepackage{times}
\usepackage{amssymb}

\newcommand{\pder}[2]{\ensuremath{\frac{\partial #1}{\partial #2}}}

\title[Supersonic Turbulence: Grid vs. SPH]{A comparison between grid and particle methods on the statistics of driven, supersonic, isothermal turbulence}

\author[Price \& Federrath]{Daniel J. Price$^{1}$, Christoph Federrath$^{2,3,1}$ \\
$^1$Centre for Stellar and Planetary Astrophysics, School of Mathematical Sciences, Monash University, Clayton Vic 3168, Australia\\
$^2$Zentrum f\"ur Astronomie der Universit\"at Heidelberg, Institut f\"ur Theoretische Astrophysik, Albert-Ueberle-Str.~2, D-69120 Heidelberg, Germany\\
$^3$Max-Planck-Institute for Astronomy, K\"onigstuhl 17, D-69117 Heidelberg, Germany
}
\date{Submitted: 28th September 2009 Revised:  20th Jan 2010 Accepted: 8th April 2010}
\pagerange{\pageref{firstpage}--\pageref{lastpage}} \pubyear{2010}

\begin{document}
\label{firstpage}
\bibliographystyle{mn2e}
\maketitle

\begin{abstract}
 We compare the statistics of driven, supersonic turbulence at high Mach number using \textsc{flash} a widely used Eulerian grid-based code and \textsc{phantom}, a Lagrangian smoothed particle hydrodynamics (SPH) code at resolutions of up to $512^{3}$ in both grid cells \emph{and} SPH particles. We find excellent agreement between codes on the basic statistical properties: a slope of $k^{-1.95}$ in the velocity power spectrum for hydrodynamic, Mach~10 turbulence, evidence in both codes for a Kolmogorov-like slope of $k^{-5/3}$ in the variable $\rho^{1/3}{\bf v}$ as suggested by \citet{kritsuketal07} and a log-normal PDF with a width that scales with Mach number and proportionality constant $b=0.33-0.5$ in the density variance--Mach number relation. The measured structure function slopes are not converged in either code at 512$^{3}$ elements.

We find that, for measuring volumetric statistics such as the power spectrum slope and structure function scaling, SPH and grid codes give roughly comparable results when the number of SPH particles is approximately equal to the number of grid cells. In particular, to accurately measure the power spectrum slope in the inertial range, in the absence of sub-grid turbulence models, requires at least $512^{3}$ computational elements in either code. On the other hand the SPH code was found to be better at resolving dense structures, giving maximum densities at a resolution of $128^{3}$ particles that were similar to the maximum densities resolved in the grid code at $512^{3}$ cells, reflected also in the high density tail of the PDF. We find SPH to be more dissipative at comparable numbers of computational elements in statistics of the velocity field, but correspondingly less dissipative than the grid code in the statistics of density weighted quantities such as $\rho^{1/3}{\bf v}$.

For SPH simulations of high Mach number turbulence we find it important to use sufficient non-linear $\beta$-viscosity in order to prevent particle interpenetration in shocks (we require $\beta_{visc} = 4$ instead of the widely used default value, $\beta_{visc} = 2$). 

\end{abstract}


\section{Introduction}
\label{sec:intro}
Dense interstellar molecular clouds are ubiquitously observed to have non-thermal line widths implying supersonic internal motions \citep{ze74}. Furthermore the amplitude of such motions increases with spatial scale in a manner reminiscent of turbulent flows in the laboratory \citep{larson81,solomonetal87,hb04}. Understanding the nature and origin of such `supersonic turbulence'  is therefore key --- perhaps \emph{the} key --- to understanding star formation \citep{es04,mk04,mo07}.  Turbulence provides a natural explanation for the clustered and hierarchical nature of star formation \citep{ElmegreenFalgarone1996}; the measured fractal dimension of interstellar gas \citep[][and references therein]{kritsuketal07,FederrathKlessenSchmidt2009}; the few percent efficiency with which gas is converted into stars \citep*{padoan95,vsbpk03,km05,elmegreen08}; most likely determines in large part the mass distribution of star forming cores (the core mass distribution, CMD) \citep{bpetal06,dibetal08} and possibly the mass distribution of stars themselves, i.e., the Initial Mass Function (IMF) \citep{pnj97,pn02,HennebelleChabrier2008}. However, given that a full theory of turbulence is elusive even in the incompressible regime apart from the phenomenology provided by \citet{kolmogorov41}, one inevitably turns to numerical simulations to glean insight. 

 Given the importance of numerical simulations in understanding the basic statistics of supersonic turbulence, and the possible implications for star formation theory, it is crucial that results inferred from such simulations are robust with respect to different numerical methods and codes. This has motivated at least two major code comparison projects in the last year or so in which both of the present authors have been involved. The `Potsdam comparison' \citep{kitsionasetal09} compared simulations of decaying, hydrodynamic turbulence using 7 different codes (3 SPH codes and 4 grid-based codes) at a fixed resolution ($215^{3}$ particles for the SPH codes, and $256^{3}$ grid cells for the grid codes). The results showed generally good agreement on statistics such as the density PDFs and power spectra, similar to earlier studies \citep{maclowetal98,khm00}. In a similar spirit the KITP07 comparison\footnote{http://kitpstarformation07.wikispaces.com/Star+Formation+Test+Problems} compared a large number of grid based and SPH calculations of decaying turbulence, for both hydrodynamics and MHD, at a range of resolutions, though the results are yet to be published.

  Both of these comparisons are problematic in several respects. The first is that it is difficult to make a statistical comparison using decaying turbulence, since the time evolution is limited and therefore only a few instantaneous snapshots can be compared. Instantaneous snapshots however are subject to intermittent fluctuations that make a head-to-head comparison based on single time slices difficult \citep{kritsuketal07,FederrathDuvalKlessenSchmidtMacLow2009}. Secondly, both comparisons start from evolved initial conditions, produced from either a previously driven SPH simulation (Potsdam) or a grid-based calculation (KITP) that has to be interpolated and/or downsampled to/from the grid/particles appropriate to the different codes, with an ensuing loss of accuracy and consistency before the comparison has even begun (this problem is much worse for the MHD case where differences in divergence-free representations for the magnetic field between codes is a further issue).

   In this paper, we consider only two codes, an SPH code, \textsc{phantom}, and a grid-based code, \textsc{flash}, which we take to be broadly representative of the fundamentally different classes of code used for star formation studies (the codes are described in \S\ref{sec:numerics}). The turbulence in both codes is driven from stationary, uniform initial conditions with exactly the same energy input and driving pattern over multiple turbulent crossing times. We also consider a range of resolutions ($128^{3}$, $256^{3}$ and $512^{3}$ in both grid elements and the number of SPH particles) in order to estimate resolution requirements and establish where convergence has occurred in one code or the other, or neither. In the present work we limit ourselves to a study of hydrodynamic turbulence, that is, without magnetic fields. This is primarily because the algorithms for Magnetohydrodynamics in SPH currently being used for star formation studies \citep[e.g.][]{pb08} rely on the Euler potentials formulation of the magnetic field, that cannot be used for turbulence studies over multiple crossing times due to the restricted field representations (see \citealt{price10} for recent progress).
 
The goals of this paper are to: i) establish whether or not agreement can be found between SPH and grid codes on the basic statistics of supersonic turbulence; ii) define resolution criteria for various statistical measures of supersonic turbulence such as power spectra, PDFs and structure functions; and iii) establish the relative strengths and weaknesses of each method for turbulence studies. We discuss the numerical methods in \S\ref{sec:numerics}, with the Fourier space driving discussed in \S\ref{sec:driving}. Results from both codes are presented in \S\ref{sec:results} and our findings discussed in \S\ref{sec:discussion}.

\section{Numerical Method}
\label{sec:numerics}
\subsection{Equations}
We solve the equations of non-self-gravitating hydrodynamics given by
\begin{eqnarray}
\pder{\rho}{t} + ({\bf v}\cdot\nabla)\rho & = & -\rho \nabla\cdot{\bf v}, \label{eq:cty} \\
\pder{{\bf v}}{t} + ({\bf v}\cdot\nabla){\bf v} & = & -\frac{\nabla P}{\rho} + {\bf f}_{stir}, \label{eq:mom}
\end{eqnarray}
where ${\bf f}_{stir}$ is a stirring force, the details of which are discussed below (\S\ref{sec:driving}). The pressure is related directly to the density via an isothermal equation of state
\begin{equation}
P = c_{s}^{2} \rho,
\label{eq:eos}
\end{equation}
where, since the equations are scale-free to all but the Mach number, we use $c_{s}=1$. We solve (\ref{eq:cty})-(\ref{eq:eos}) using periodic boundary conditions in the three-dimensional domain $x,y,z \in [0,1]$. The initial conditions are a uniform density medium $\rho = \rho_{0} = 1$ with zero initial velocities.

 We discuss our results in terms of the dynamical time, defined as $t_{d} \equiv L/(2\mathcal{M})$, where $L$ is the box size and $\mathcal{M}$ is the RMS Mach number.
However, the absence of a physical scale in the equations means that the results can be arbitrarily scaled to the interstellar medium by defining length, mass and time scales. For example, adopting a length scale of 10~pc and a sound speed of 0.2~km/s, gives the physical time from $t/t_{d}$ according to
\begin{equation}
t_{physical} = 2.5~\textrm{Myr} \left(\frac{L}{10~\textrm{pc}}\right) \left(\frac{c_{s}}{0.2~\textrm{km/s}} \right)^{-1} \frac{t}{t_{d}}.
\end{equation}
One can similarly set a mass scale by defining the initial density to be $n_{0}\approx 3 \times 10^{2}$~cm$^{-3}$, i.e., $\rho_{0} \approx 10^{-21}~\mathrm{g}\,\mathrm{cm}^{-3}$ assuming fully molecular hydrogen gas, giving a total mass in the box of $\sim 1.5 \times 10^{4}~{\rm M}_{\odot}$. For these parameters, the maximum density reached in our calculations from turbulent fluctuations alone is $\rho_{max} \sim 10^{-17}$--$10^{-16}\,\mathrm{g}\,\mathrm{cm}^{-3}$ or $n_{max}\sim 10^{6}$ cm$^{-3}$ (see Fig.~\ref{fig:rhomax}) for the highest resolution SPH calculation.

\subsection{Driving}
\label{sec:driving}
For driving turbulence we use the same driving routine used in\citet{SchmidtHillebrandtNiemeyer2006,FederrathKlessenSchmidt2008,FederrathDuvalKlessenSchmidtMacLow2009,FederrathKlessenSchmidt2009} and \citet{SchmidtEtAl2009}. The driving routine updates a vector of real values according to an algorithm that generates an Ornstein-Uhlenbeck, or ``coloured noise" sequence \citep[e.g.][]{EswaranPope1988}. The sequence $x_{n}$ is a Markov process that takes the previous value, weights by an exponential damping factor with a given auto-correlation time $t_{s}$, and drives by adding a Gaussian random variable, weighted by a second damping factor, also with correlation time $t_s$. For a timestep $dt$, this sequence can be written as:
\begin{equation}
     x_{n+1} = f x_{n} + \sigma \sqrt{(1 - f^{2})}\,z_{n}
\end{equation}
where $f = \exp (-dt / t_{s})$, and $z_n$ is a Gaussian random variable drawn from a Gaussian distribution with unit variance, and $\sigma$ is the desired variance of the Ornstein-Uhlenbeck sequence \citep[see e.g.][]{bartosch01}. The resulting sequence satisfies the properties of zero mean, and stationary RMS equal to $\sigma$. Its power spectrum in the time domain can vary from white noise ($P(f) = \mathrm{const}$) to ``brown" noise ($P(f) \propto 1 / f^2$).


The physical forcefield is constructed in Fourier space using the Ornstein-Uhlenbeck process. This allows a simple decomposition of the field into a solenoidal (divergence-free) part and a compressible (curl-free) part using a Helmholtz decomposition. In this study, we only keep the solenoidal part. Inverse Fourier transformation yields the physical solenoidal force field ${\bf f}_{stir}$ used in equation (\ref{eq:mom}). A more detailed description of the forcing module applied here is provided in \citet{FederrathDuvalKlessenSchmidtMacLow2009}.

The time-dependent Fourier modes for constructing the forcing patterns ${\bf f}_{stir}$ were calculated and written to a file before the actual numerical experiments. Both the SPH and the grid code read exactly the same forcing sequence from this file. Thus, it was guaranteed that both codes were using exactly the same forcing at all times during the comparison experiments.

\subsection{{\sc flash} (grid)}
\subsubsection{Hydrodynamics}
\textsc{flash} \citep{FryxellEtAl2000,DubeyEtAl2008} is an adaptive-mesh refinement code \citep{BergerColella1989} that uses the piecewise parabolic method \citep[PPM,][]{ColellaWoodward1984} to solve the equations of hydrodynamics. The PPM provides a shock capturing scheme to keep shocks and contact discontinuities sharp (typically spreading over 2-3 zones), while maintaining third order accuracy in smooth flows through a parabolic reconstruction scheme. In this study, \textsc{flash} v3 was used, which provides a uniform grid mode. Thus, the overhead in storing and iterating the adaptive mesh hierarchy was completely removed, which yields a speed-up of factors of a few. \textsc{flash} is parallelised using the message passing interface (MPI). For the resolutions studied here ($128^3$, $256^3$ and $512^3$ grid cells), 1, 8 and 64 MPI processes respectively were used in a mode of parallel computation, each calculation taking roughly 12, 250 and 5000 CPU-hours respectively. \textsc{flash} has been extensively tested against laboratory experiments \citep{CalderEtAl2002} and other codes \citep{DimonteEtAl2004,HeitmannEtAl2005,kitsionasetal09}.

\subsubsection{Tracer particles}
\label{sec:tracers}
\textsc{flash} provides an option for Lagrangian tracer particles, which can be evolved alongside the hydrodynamics. Similar to SPH particles, tracer particles provide information in the Lagrangian frame, but unlike SPH particles, the tracer particles have no feedback on the hydrodynamics, i.e. the variables on the grid are independent of the tracers. The tracer particles' $x$, $y$ and $z$ positions can be any real number within the computational domain, not bound to the grid. However, they are moved with the velocity computed on the grid. The velocity is interpolated at the exact position of each tracer particle for each timestep using a first order cloud-in-cell interpolation scheme. Higher-order interpolation schemes like the triangular-shaped-cloud scheme can also be used instead. However, we used the first-order scheme here, because various tests suggested no strong dependence of our results on the interpolation scheme. The tracer particles were moved on the hydrodynamic timestep with the grid-interpolated velocity using a first-order scheme. We initialised $128^3$, $256^3$ and $512^3$ tracer particles at $t=0$ on a uniform grid at exactly the same positions as the SPH particles were initialised in the \textsc{phantom} calculations (see~\S\ref{sec:sphinit}), matching the grid and SPH resolutions ($128^3$, $256^3$ and $512^3$, respectively). Adding the tracer particles does not add any significant computational overhead to the \textsc{flash} calculations, apart from the additional memory requirements.

 In order to extract the maximum possible information from the tracer particles, we have computed --- in post-processing --- a density field based solely on the tracer particle positions. This is achieved by assuming they are particles of fixed mass (dividing the total mass in the simulation by the number of tracer particles) and using the SPH density calculation routine from \textsc{phantom} where the density and smoothing length are iterated self-consistently (based on Eqs.~\ref{eq:rhosum} and~\ref{eq:hrho}). Column-integrated and cross-section slice plots of the density field were then produced as for the \textsc{phantom} results using \textsc{splash} \citep{splashpaper}.

\subsection{{\sc phantom} (SPH)}
\textsc{phantom} is a low-memory, highly efficient SPH code written especially for studying non-self-gravitating problems. The code is made very efficient by using a simple neighbour finding scheme based on a fixed grid and linked lists of particles. The calculations shown in this paper have used only the shared memory \textsc{openMP} parallelisation in \textsc{phantom}, using 4, 8 and 32 processors and requiring 265, 5050 and 120,000 CPU-hours for the $128^{3}$, $256^{3}$ and $512^{3}$ calculations, respectively. Thus the $256^{3}$ \textsc{phantom} calculation was roughly comparable in computational cost to the $512^{3}$ \textsc{flash} calculation, and similarly for the $128^{3}$ \textsc{phantom} vs. $256^{3}$ \textsc{flash} (though some caution is required here due to the different machines and architectures used to run each code). One may also consider that \textsc{phantom} was found to be roughly an order of magnitude faster than `standard' SPH codes in the \citet{kitsionasetal09} turbulence comparison.

\subsubsection{Hydrodynamics}
\label{sec:sphhydro}
 For hydrodynamics \textsc{phantom} implements the full variable smoothing length SPH formulation developed by \citet{pm04b} and \citet{pm07}, whereby the smoothing length, $h$, and density, $\rho$, are mutually dependent via the density sum (for particle $a$)
\begin{equation}
\rho_{a} = \sum_{b} m_{b} W_{ab} (h_{a}),
\label{eq:rhosum}
\end{equation}
which is an exact solution to (\ref{eq:cty}), and the relation
\begin{equation}
h_{a} = \eta \left(\frac{m_{a}}{\rho_{a}}\right)^{1/3},
\label{eq:hrho}
\end{equation}
where $m$ is the particle mass and $W_{ab} \equiv W(\vert {\bf r}_{a} -  {\bf r}_{b}\vert, h_{a})$ is the SPH smoothing kernel (see e.g. \citealt{monaghan92,price04,monaghan05} for reviews of SPH). Equations (\ref{eq:rhosum}) and (\ref{eq:hrho}) are iterated self-consistently using a Newton-Raphson method as described in \citet{pm07}, where in this paper we have used $\eta=1.2$, giving approximately 58 neighbours per particle in a smooth distribution.

 The fact that the smoothing length has a functional dependence on (ultimately) the particle position means that the derivatives of $h$ can be accounted for in the equations of motion, resulting in exact conservation of momentum, angular momentum, energy and entropy in the SPH equations. In \textsc{phantom} the equations of motion (\ref{eq:mom}) take the form
\begin{eqnarray}
\frac{d{\bf v}_{a}}{dt} & = & -\sum_{b} m_{b} \left[ \frac{P_{a} + q_{a}}{\Omega_{a}\rho_{a}^{2}} \nabla_{a} W_{ab} (h_{a})\right. \nonumber \\
&  & \phantom{\sum_{b} m_{b} [} +\left. \frac{P_{b} + q_{b}}{\Omega_{b}\rho_{b}^{2}} \nabla_{a} W_{ab} (h_{b})\right] + {\bf f}_{stir},
\label{eq:sphmom}
\end{eqnarray}
where $P$ is the pressure, $\Omega$ is a dimensionless quantity related to the smoothing length gradients (see \citealt{pm07} for details) and $q$ represents the artificial viscosity term (discussed below). In the absence of shock dissipation ($q=0$) there is zero numerical dissipation contained in the above equations and energy is conserved to the accuracy of the timestepping scheme  --- here a Kick-Drift-Kick leapfrog integrator equivalent to the velocity Verlet method, implemented with individual particle timesteps. For an isothermal equation of state the viscosity term  $q$ therefore represents the only numerical form of energy loss.

\subsubsection{Artificial viscosity}
\label{sec:av}
 Shocks are treated in \textsc{phantom} using a standard artificial viscosity term, though formulated slightly differently to the usual SPH expression in order to obtain a more efficient calculation. Instead of the usual expression we write the artificial viscosity as $q$ in (\ref{eq:sphmom}), where
\begin{equation}
q_{a} = \left\{
\begin{array}{ll}
\frac12 \alpha_{a} \rho_{a} v_{sig,a} \vert {\bf v}_{ab}\cdot\hat{\bf r}_{ab} \vert,& {\bf v}_{ab}\cdot\hat{\bf r}_{ab} < 0 \\
0 &  {\bf v}_{ab}\cdot\hat{\bf r}_{ab} \ge 0
\end{array}\right.
\label{eq:qvisc}
\end{equation}
where ${\bf v}_{ab} \equiv {\bf v}_{a} - {\bf v}_{b}$ and we use
\begin{equation}
v_{sig,a} = c_{s,a} + \beta_{visc} \vert {\bf v}_{ab}\cdot\hat{\bf r}_{ab} \vert
\label{eq:vsig}
\end{equation}
as the maximum signal velocity for hydrodynamics. The $\beta$ term in the signal velocity provides a non-linear term that was originally introduced to prevent particle penetration in high Mach number shocks (see e.g. \citealt{monaghan89}). Indeed one of our findings from this comparison is that sufficient $\beta$-viscosity is an important factor for accurate SPH calculations in the supersonic regime. We use $\beta_{visc} = 4$ in this paper, the motivation for which is discussed further in \S\ref{sec:coldens} and demonstrated in Appendix~\ref{sec:viscosity}.

 The artificial viscosity described above is essentially the same as the \citet{monaghan97} formulation with a slightly different averaging of the density and signal velocity. We use the \citet{mm97} switch to reduce dissipation away from shocks, in which the dissipation parameter $\alpha$ is evolved according to a source and decay equation
\begin{equation}
\frac{d\alpha_{a}}{dt} = - \frac{\alpha_{a} - \alpha_{min}}{\tau_{a}} + \mathcal{S}_{a}, \hspace{1cm} \tau_{a} = h_{a}/(\sigma c_{s})
\end{equation}
where we have used $\sigma=0.1$, $\mathcal{S} = \max (0,-\nabla\cdot{\bf v})$, $\alpha_{min} = 0.05$, enforced $\alpha_{max}=1.0$, and given all particles $\alpha = \alpha_{min}$ initially.

\subsubsection{Boundary conditions}
Periodic boundary conditions are implemented in \textsc{phantom} by directly finding neighbours across the periodic boundary and interacting with a distance in each direction calculated according to
\begin{equation}
(x_{a} - x_{b}) = \min \left[(x_{a} - x_{b}), \left(\vert x_{a} - x_{b}\vert - L\right) \frac{x_{a} - x_{b}}{\vert x_{a} - x_{b}\vert} \right],
\end{equation}
where $L$ is the box size in the corresponding direction. This gives a significant memory saving since memory does not have to be assigned to the storage of ghost particles.

\subsubsection{Initial conditions} \label{sec:sphinit}
The SPH particles were set up initially on a regular cubic lattice, using equal mass particles, identical to the setup used for the Lagrangian tracer particles in the \textsc{flash} calculations (\S\ref{sec:tracers}).

\begin{figure}
\begin{center}
 \includegraphics[width=\columnwidth]{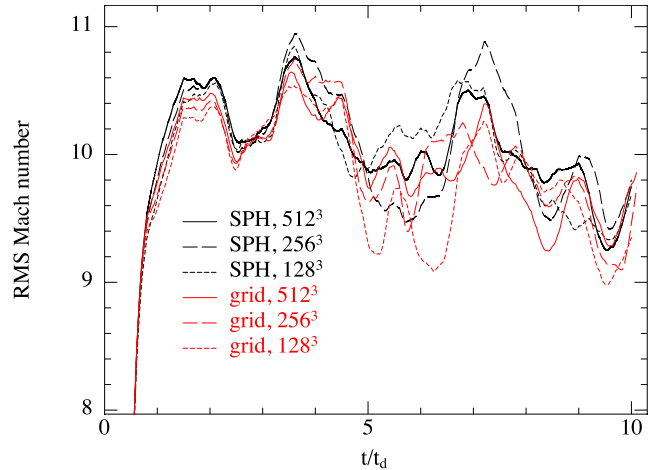}
 \caption{Mass-weighted RMS Mach number as a function of time for the six calculations, as indicated in the legend. The time evolution is similar for both the SPH and the grid code and for all resolutions up to $\sim 4 t_{d}$, where all calculations show deviations from each other of order 5\% in $\delta \mathcal{M}/\mathcal{M}$ though with no systematic trends with either code or resolution.}
\label{fig:vrms}
\end{center}
\end{figure}

\begin{figure}
\begin{center}
 \includegraphics[width=\columnwidth]{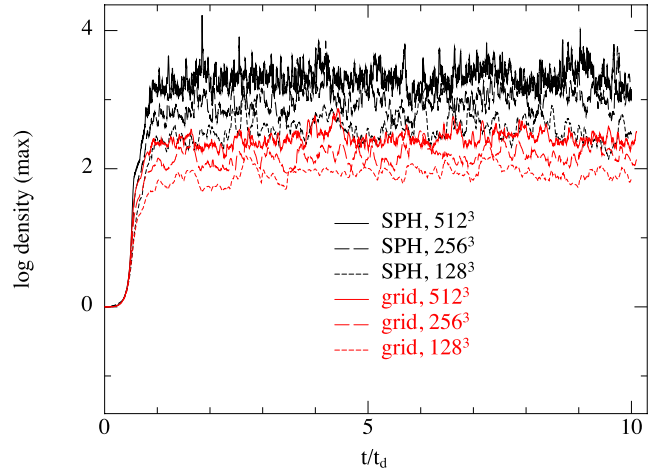}
 \caption{Maximum density as a function of time in the calculations using \textsc{phantom}  (SPH, black lines) and \textsc{flash} (grid, red lines) at resolutions of $128^{3}$, $256^{3}$ and $512^{3}$ particles/grid cells, as indicated. The evolution shows strong time variability, a consequence of the intermittency inherent in the log-normal probability distribution function (increasingly higher densities have correspondingly smaller probabilities). The maximum density is therefore also a strong function of resolution in each code. At $128^{3}$ particles the SPH code resolves maximum densities similar to those achieved at $512^{3}$ on the grid.}
\label{fig:rhomax}
\end{center}
\end{figure}

\section{Results}
\label{sec:results}
 Both codes have been run using the same driving pattern for 10 dynamical times ($t=0.5$ in code units) and at a resolution of $128^{3}$, $256^{3}$ and $512^{3}$ elements. For the grid code, this resolution is fixed spatially throughout the evolution, giving fixed resolution in volume but variable resolution in mass, whilst for the SPH code the particles move following the fluid motion, giving equal resolution in mass but variable resolution in volume. The two methods are therefore very nicely complementary for assessing the statistics of supersonic turbulence for different quantities which may be either mass or volume-weighted. It should be noted that whilst grid-based calculations of ISM turbulence have been run at much higher resolutions --- up to $2048^{3}$, see \citealt{kritsuketal07}, with resolutions of $1024^{3}$ even in early simulations of decaying turbulence \citep{pwp98} ---, our use of 134,217,728 SPH particles ($512^{3}$) represents the highest resolution turbulence simulation performed to date with an SPH code, over an order of magnitude higher than the ``high resolution'' calculation (10 million particles) in \citet{bpetal06} and two-and-a-half orders of magnitude higher than the $\sim$200,000 particles used for many of the runs in that paper and elsewhere \citep{khm00,vsbpk03}.

\subsection{Time evolution of global variables}
 The time variation of the mass-weighted RMS Mach number and of the maximum gas density are shown in Figures~\ref{fig:vrms} and \ref{fig:rhomax}, computed at every timestep for both codes. The mass-weighted RMS Mach number in SPH is simply the square root of the average value of $v^{2}/c_{s}^{2}$ on the particles, whilst in \textsc{flash} this has been computed using the RMS value of $\rho v^{2}/(\rho_{0} c_{s}^{2})$, to correspond to the SPH average.
 
\begin{figure*}
\begin{center}
 \includegraphics[width=0.8\textwidth]{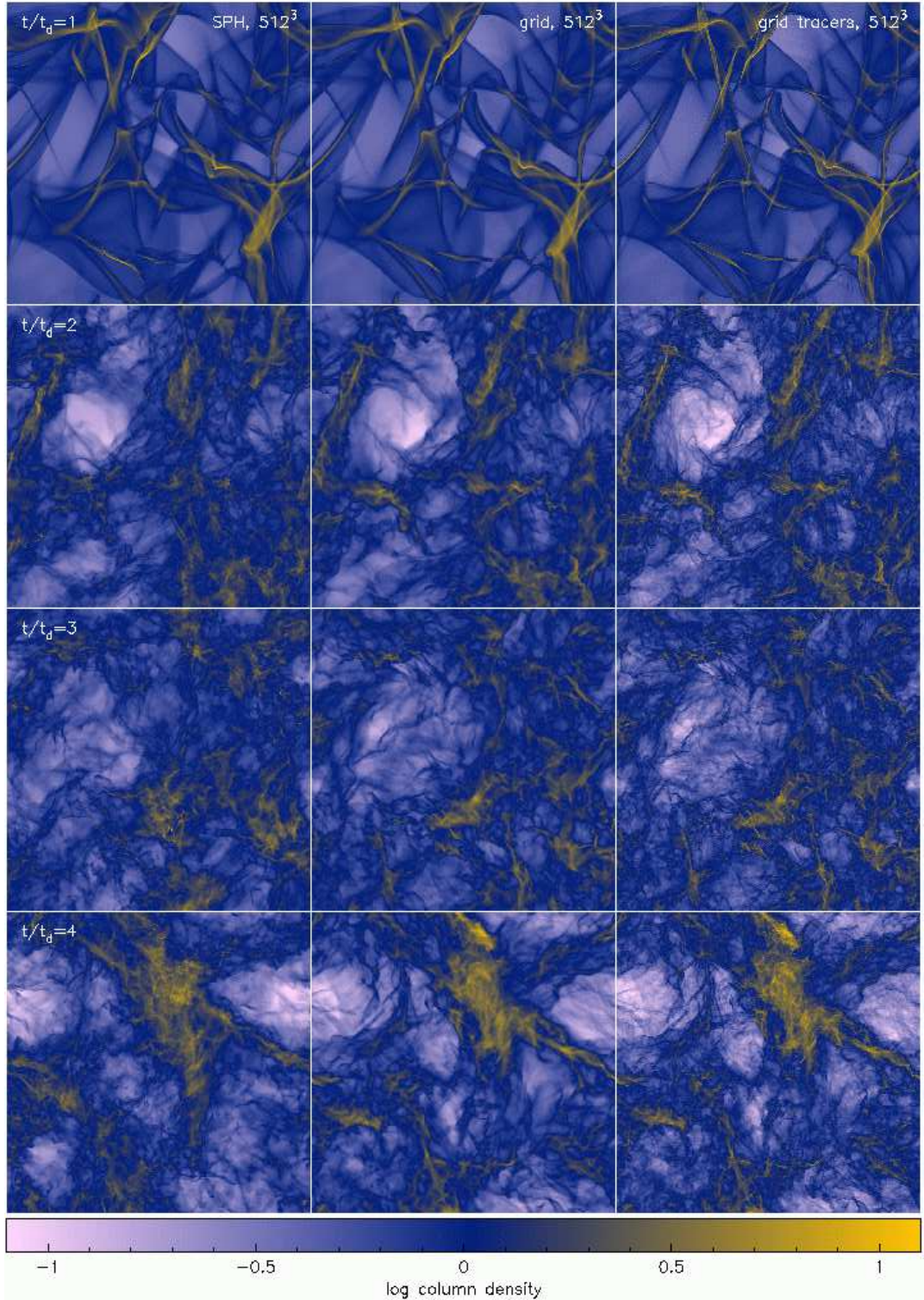}
 \caption{Projected column density in the \textsc{phantom} (SPH, left) and \textsc{flash} (grid, centre) calculations at a resolution of $512^{3}$ particles/grid cells, showing the evolution over the first few dynamical times (top to bottom), together with the density computed from the tracer particle positions in the \textsc{flash} calculation using an SPH density estimate (right panel). After 1 dynamical time (top row), there is clear correspondence in individual shock structures between the SPH and the grid code, whilst after 2 dynamical times (second row) there are similar large scale features. However by 3 or 4 dynamical times (third and fourth row) only a weak correlation even between large scale features is observed. Dense features are in general better resolved in the SPH calculations at equivalent resolutions (in a number of particles=number of grid cells sense), whilst the grid-based calculations tend to better resolve features in low density regions (see also Fig.~\ref{fig:slicerho_t10}). The increased resolution of sharp features in the tracer particle density fields (right column, compared to centre column) suggest a remarkable ability for the tracer particles to provide information on sub-grid scales at essentially zero additional computational cost.}
\label{fig:coldens512}
\end{center}
\end{figure*}

  The Mach number evolution (Fig.~\ref{fig:vrms}) is similar in both codes and at all resolutions up to around 4 $t_{d}$, at which point all calculations show variations of order 5\% from each other. No clear trends either between codes or with resolution are apparent, indicating that the variation observed is due to the stochastic nature of fully-developed turbulence producing different though statistically similar evolution (see Fig.~\ref{fig:coldens512} for the evidence for this in the density field). It is these intermittent fluctuations that make turbulence code comparisons based on snapshot-to-snapshot comparison difficult, because the instantaneous turbulence field will quickly diverge between different codes in the fully developed regime because of the chaotic nature of the turbulence (see projections and slices for $t \gtrsim 2 t_{d}$ comparing the SPH and grid results).

  
  The evolution of maximum density (Fig.~\ref{fig:rhomax}) shows strong time variability in all six calculations, similar to the results shown in \citet[][Fig.~2]{kritsuketal07} and \citet[][Fig.~2]{FederrathKlessenSchmidt2009}. For isothermal flows arbitrarily large density fluctuations can be produced, though with vanishingly small probability as can be inferred from the log-normal form of the PDF. This simply reflects the highly intermittent nature of the density fluctuations in supersonic, turbulent flows. The maximum density is a clear function of resolution in each code, showing no signs of convergence, as one might expect seeing as we are sampling the very highest data point in the PDF. The results also demonstrate the evidently higher mass resolution in the SPH code: at $128^{3}$ particles the maximum density resolvable in SPH is roughly similar to that resolved at $512^{3}$ on the grid. Using $512^{3}$ SPH particles the maximum density resolved at RMS Mach 10 is roughly three-and-a-half orders of magnitude above the mean density which one might therefore expect to be similar to the mass resolution in a $2048^{3}$ grid-based calculation.

\begin{figure*}
\begin{center}
\includegraphics[width=0.9\textwidth]{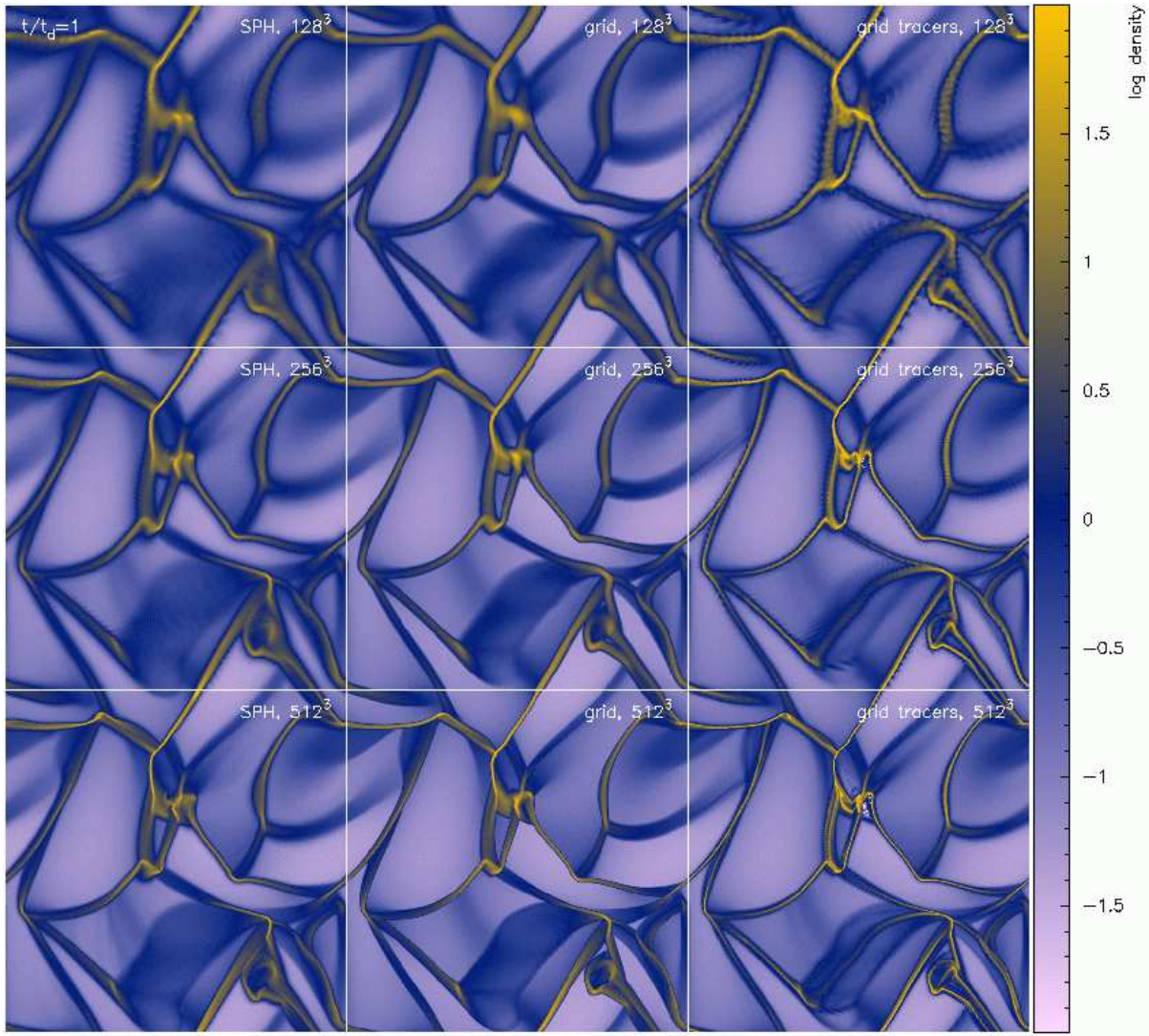}
 \caption{Cross section slice of the density at the box midplane ($z=0.5$) after 1 dynamical time, for three different resolutions ($128^{3}$, $256^{3}$ and $512^{3}$ in grid cells/particles, top to bottom) using \textsc{phantom} (SPH, left), \textsc{flash} (grid, centre) and for the density calculated from the tracer particles in the grid calculation using an SPH summation (right panels).}
\label{fig:slicerho_t1}
\end{center}
\end{figure*}

\begin{figure*}
\begin{center}
 \includegraphics[angle=270,width=0.9\textwidth]{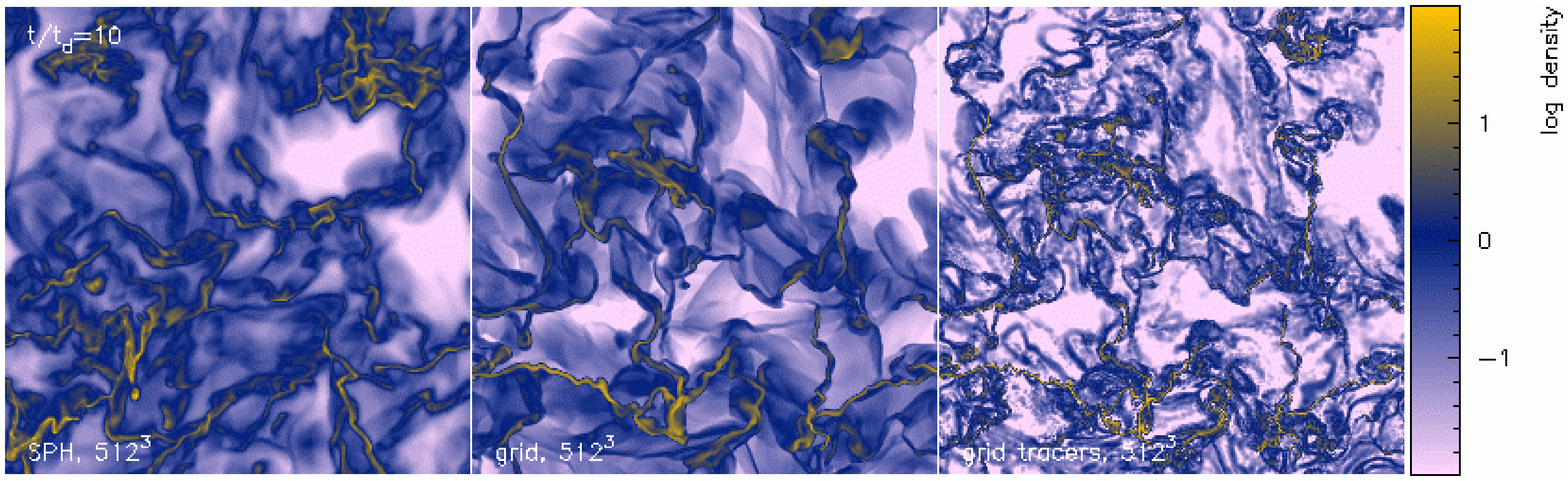}
\caption{Cross section slice of the density at the box midplane ($z=0.5$), as in Fig.~\ref{fig:slicerho_t1} but here shown after 10 dynamical times and showing only the highest resolution calculations ($512^{3}$). The \textsc{flash} calculations (grid, centre) shows better resolution in low density regions compared to \textsc{phantom} (SPH, left). In evolved snapshots the tracer particles appear strongly clustered in high density regions and almost completely absent from the voids (right panel).}
\label{fig:slicerho_t10}
\end{center}
\end{figure*}


\subsection{Density field}

\subsubsection{Projected density fields}
\label{sec:coldens}
 The projected column density fields at the highest resolution ($512^{3}$) are shown for \textsc{phantom}, \textsc{flash} and the density field computed from the \textsc{flash} tracer particles in Fig.~\ref{fig:coldens512} (left, middle and right columns, respectively), at intervals of $\Delta t= 1 t_{d}$ for the first four dynamical times. The column density plots for SPH have been produced directly from the particles to a 2D pixel map using the \textsc{splash} visualisation tool \citep{splashpaper} whilst the grid based results have been integrated through the grid. We show the integration through the $z$-direction in the codes.

  At early times the calculations show clear agreement in the location of individual shocks ($t=1 t_{d}$, top row) and in the development of large scale structures ($t=2 t_{d}$, second row). By $t=4  t_{d}$ (fourth row) there is no longer clear correspondence even at the largest scales between codes, in agreement with the observed deviations in the time evolution of the RMS Mach number around this time (Fig.~\ref{fig:vrms}).


 In terms of resolution, high-density structures appear better resolved in the SPH calculations at the same number of computational elements. However, the grid results tend to show better resolution of features in low density regions, as one might expect since in SPH the resolution is preferentially shifted \emph{away} from low density regions towards high density regions.
 
  The excellent agreement between codes in the development of individual shock structures within the first dynamical times (top row) enabled us to make a very detailed comparison of features between codes which proved to be very helpful in the comparison process. In particular it highlighted that, with the parameters we were initially using, some of the dense structures created by the collision of one or more shocks were rapidly losing definition in the SPH results, resulting in a noisy density field that was rather unlike the grid results (this is shown in more detail in Appendix~\ref{sec:viscosity}). The problem could be easily traced to be caused by particles penetrating or ``overshooting'' the shock front in these high Mach number shocks, a problem which the non-linear $\beta$ (von-Neumann-Richtmyer) term in the SPH artificial viscosity (Eqs. \ref{eq:qvisc}-\ref{eq:vsig}) was designed to prevent \citep[see][]{monaghan89}. The problem was thus easily fixed by using a larger value for $\beta_{visc}$. We have therefore used $\beta_{visc} = 4$ throughout the paper, rather than the nominal $\beta_{visc} = 2$ which is widely used --- and sufficient --- for low Mach number calculations. It should be noted that this makes very little difference to the overall dissipation rate since the linear viscosity term ($\alpha$) dominates the numerical dissipation rate almost everywhere except at very strong divergence in the velocity field (where particle penetration can occur).

 Comparing the projected column density fields calculated using the tracer particles in the \textsc{flash} calculation (right column) to the grid-based density field (centre column), showing a wealth of sub-grid structures, suggests that the tracer particles have the ability to provide a truly staggering improvement in resolution in the density field. The improved resolution is all the more remarkable considering that the tracer particles are merely advected with the grid-based velocity field at essentially no extra computational expense.

\subsubsection{Cross section slices}
\label{sec:slices}
\begin{figure}
\begin{center}
 \includegraphics[width=\columnwidth]{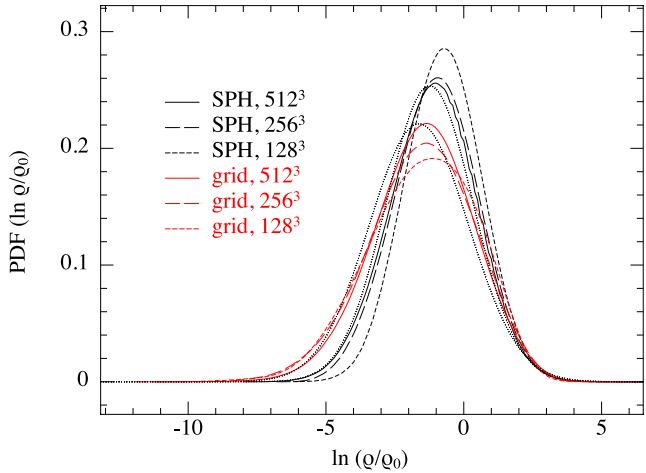}
 \caption{Time-averaged Probability Distribution Function (PDF) of the logarithm of the density field $s\equiv \ln \rho$ from the \textsc{phantom} (SPH, black lines other than dotted) and \textsc{flash} (grid, red lines) calculations, each at resolutions of $128^{3}$, $256^{3}$ and $512^{3}$ particles/grid cells. The PDFs are averaged over 81 snapshots evenly spaced between $t/t_{d}=2$ and $t/t_{d}=10$. Here we show the PDFs on a linear scale to highlight the change in position of the peak value as a function of resolution. We have also plotted, as dotted black lines, the best fit (to the peak) log-normal distributions from equations (\ref{eq:lognormal}) and (\ref{eq:bfac}) using $b=0.5$ (best fitting the $512^{3}$ grid results) and $b=0.33$ (best fitting the $512^{3}$ SPH results). The SPH and grid results show complementary trends with resolution.}
\label{fig:pdflin}
\end{center}
\end{figure}

\begin{figure}
\begin{center}
\includegraphics[width=\columnwidth]{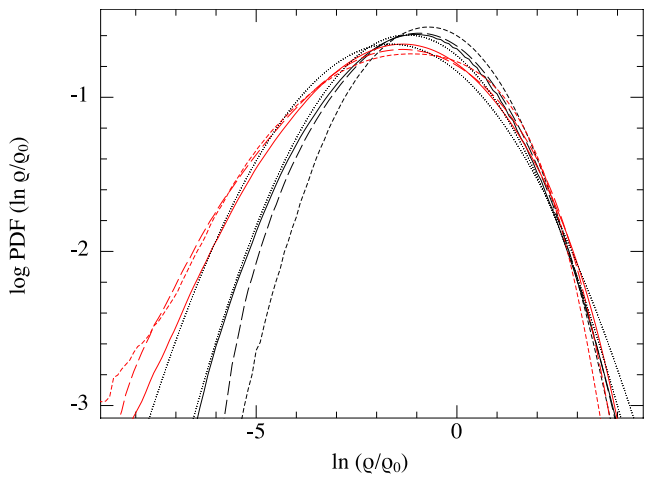}
 \caption{Probability Distribution Function (PDF) of $\ln \rho$, as in Fig.~\ref{fig:pdflin}, but here shown on a logarithmic scale. The PDFs are log-normal to good approximation for $\sim 3-4$ orders of magnitude in density either side of the mean, demonstrated by the best fit log-normal distributions (fitted to the peak in Fig.~\ref{fig:pdflin}) given by the dotted black lines using $b=0.5$ and $b=0.33$ to fit the $512^{3}$ grid (solid red line) and $512^{3}$ SPH (solid black line) results respectively.}
\label{fig:pdflog}
\end{center}
\end{figure}

\begin{figure}
\begin{center}
 \includegraphics[width=\columnwidth]{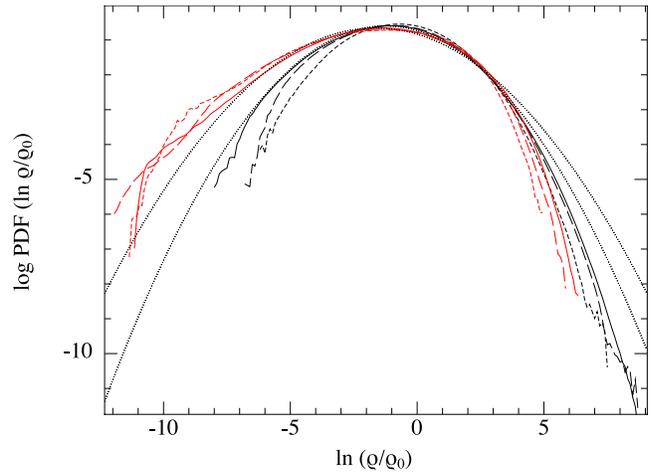}
 \caption{Time averaged PDFs, as in Figures~\ref{fig:pdflin} and \ref{fig:pdflog} but here showing the tails of the distributions at very low and very high densities. The SPH code (black lines other than dotted) resolves the PDF to much higher densities than the grid code (red lines), though the grid results correspondingly extend further into the low density regime. The log-normal distributions (dotted black lines) are no longer a good approximation to the distribution very far from the mean, in particular in the high density tail (where the $256^{3}$ and $512^{3}$ SPH results appear to show convergence).}
\label{fig:pdftails}
\end{center}
\end{figure}

 Column density plots such as those shown in Fig.~\ref{fig:coldens512} in general tend to highlight dense features, since all structures along the line of sight contribute to the projected field (which also tends to be the case in observations). The features in column density plots are therefore reflected by statistics such as the PDF (Figs.~\ref{fig:pdflin}--\ref{fig:pdftails}) and quantities such as the maximum density (Fig.~\ref{fig:rhomax}). However volumetric quantities such as the volume filling factor of the material and the velocity field, reflected in statistics such as power spectra and structure functions, are better illustrated by cross-section slices. For this reason we show cross section slices of density at the midplane of the computational domain ($z=0.5$), showing a resolution study of the initial shock development at 1 dynamical time (Fig.~\ref{fig:slicerho_t1}) and a comparison of the evolved snapshots at the end of the simulations ($t/t_{d}=10$), showing only the highest resolution (Fig.~\ref{fig:slicerho_t10}). The plots show the density field using \textsc{phantom} (left columns in Figs.~\ref{fig:slicerho_t1} and \ref{fig:slicerho_t10}), \textsc{flash} (centre column in both Figs.) and for the tracer particle density field computed from the \textsc{flash} calculations (right columns).
 
  Figs.~\ref{fig:slicerho_t1} and \ref{fig:slicerho_t10} show clearly that the grid results are better resolved in low density regions. The resolution in the SPH calculations is concentrated towards high density regions which fill relatively little of the volume. Comparing individual shock structures in Fig.~\ref{fig:slicerho_t1} shows that in general the shocks have better definition in \textsc{flash}, with the shock widths in the highest resolution \textsc{phantom} calculation similar to those obtained at $256^{3}$ in \textsc{flash}. This is as might be expected given the relative crudeness of the shock capturing scheme (artificial viscosity) in the SPH code compared to the PPM shock capturing scheme \citep{ColellaWoodward1984} employed in \textsc{flash}. In the more evolved snapshots (Fig.~\ref{fig:slicerho_t10}), the grid results show many well-defined shock features in low density regions that are much less well resolved in the SPH calculations.
  
   Some numerical artefacts are visible in the lowest resolution SPH calculations in the earliest snapshot ($t=1 t_{d}$, top left panel of Fig.~\ref{fig:slicerho_t1}) due to the ``breaking'' of the initial regular lattice on which the particles were placed as it is distorted by the flow. Interestingly similar artefacts are visible --- and more accentuated --- in the low resolution tracer particle plots (top right panel). These effects are not obviously visible either in the SPH or the tracer particles at higher resolution (middle and bottom rows of Fig.~\ref{fig:slicerho_t1}) or at later times (Fig.~\ref{fig:slicerho_t10}) once the particles have adopted a more `natural' arrangement. There are no obvious artefacts at low resolution in the grid based calculation.

  Density slices calculated from the tracer particles in the \textsc{flash} calculation are shown in the rightmost panels of Figs.~\ref{fig:slicerho_t1} and \ref{fig:slicerho_t10}, using the SPH density summation (\ref{eq:rhosum}) iterated self-consistently with the smoothing length according to (\ref{eq:hrho}). Comparison with the grid-based density field at 1 dynamical time, the tracer particles appear to substantially increase the resolution in high density regions. Whilst most of the features have close correspondence to those visible in the grid slices (centre column of Fig.~\ref{fig:slicerho_t1}), it is notable that a dense shock structure appears in the lower part of the $t/t_{d}=1$ snapshots at all resolutions that is completely absent from both the SPH and grid density fields. The absence of this feature even at $512^{3}$ in the centre column of Fig.~\ref{fig:slicerho_t1}, yet clearly present at $128^{3}$ in the tracer particles, suggests that it may be an artefact of tracer particles clustering below the grid scale. At later times (right panel of Fig.~\ref{fig:slicerho_t10}) this is even more evident by the fact that the tracer particles are strongly concentrated in high density regions and largely evacuated from low density regions (i.e., large parts of the panels are saturated at the density floor of the plot due to the absence of a contribution from tracer particles even with iterated smoothing lengths).

 The difference between the density slices and column density plots shows that in general, \emph{for similar numbers of computational elements} (not the same as equal computational expense), SPH codes are better at resolving dense structures (highlighted by projections through the volume), whilst grid codes are better at resolving volumetric structures (highlighted by slices though the volume).

\subsection{Probability Distribution Functions}
\label{sec:pdfs}
  Many studies have demonstrated that the density probability distribution function (PDF) in supersonic turbulence is well represented by a log-normal distribution \citep[e.g.][]{pnj97,pvs98,klessen00,kritsuketal07,FederrathKlessenSchmidt2008,ls08,FederrathDuvalKlessenSchmidtMacLow2009}, i.e.,
\begin{equation}
p(\ln\rho)d\ln\rho = \frac{1}{\sqrt{2\pi\sigma^{2}}} \exp{\left[ -\frac12 \left(\frac{\ln \rho - \overline{\ln \rho} }{\sigma}\right)^{2}\right]} d\ln\rho.
\label{eq:lognormal}
\end{equation}
where the mean of the logarithm of density $\ln \rho$ is related to the standard deviation $\sigma$ of $\ln \rho$ by
\begin{equation}
\overline{\ln\rho} = -\sigma^{2}/2.
\end{equation}
 The appearance of a log-normal form in isothermal flows can be understood analytically as a consequence of the multiplicative central limit theorem assuming that individual density perturbations are independent and random \citep{vs94,pvs98,np99}. In physical terms this means that density fluctuations at a given location are constructed by successive passages of shocks with a jump amplitude independent of the local density \citep{bpetal07,kritsuketal07,FederrathDuvalKlessenSchmidtMacLow2009}. Furthermore the width of the PDF for $\ln\rho$ is found to be related to the rms Mach number according to
\begin{equation}
\sigma^{2} = \ln\left(1 + b^{2}\mathcal{M}^{2} \right),
\label{eq:bfac}
\end{equation}
where the factor $b \approx 1/2$ has been suggested by early numerical experiments \citep[e.g.][]{pnj97}. More recently \citet{kritsuketal07} find a much lower value of $b\approx 0.26$ whilst \citet{beetzetal08} find $b=0.37$. \citet{brunt10} has also recently measured $b=0.5\pm 0.05$ in the Taurus Molecular Cloud, based on a method for inferring the 3D variance from 2D observations developed by \citet{bfp10a}. \citet{FederrathKlessenSchmidt2008} and \citet{FederrathDuvalKlessenSchmidtMacLow2009} reconcile these results by showing that the width of the PDF depends not only on the RMS Mach number but also on the relative degree of compressible and solenoidal modes in the forcing, with $b=0.33$ appropriate for purely solenoidal forcing and b=1 for purely compressive forcing. \citet{ls08} --- performing calculations at a range of Mach numbers --- suggest that the relationship (\ref{eq:bfac}) should be adjusted, finding $\sigma^{2} = -0.72 \ln \left(1 + 0.5 \mathcal{M}^{2} \right) + 0.20$ from a three-parameter fit for hydrodynamic turbulence.
For the purposes of the comparison at hand we simply fit the PDFs using a single parameter $b$ based on Eqs. (\ref{eq:lognormal})--(\ref{eq:bfac}).

\subsubsection{Volume weighted PDFs}
\label{sec:vwpdfs}
 Time-averaged PDFs of $s\equiv\ln(\rho/\rho_{0})$ for the three SPH calculations and three grid calculations are shown in Figures~\ref{fig:pdflin}, \ref{fig:pdflog} and \ref{fig:pdftails}. The plots show the time average of individual PDFs computed at intervals of $\Delta t = t_{d}/10$, starting from $2 t_{d}$ when turbulence is reasonably well established (see Figures~\ref{fig:vrms}, \ref{fig:coldens512}). This gives a total of 81 snapshots used in the averaging procedure. For reference we compute the PDF for each code with a bin width of 0.1 in $\ln \rho$, with the first bin starting at $(\ln\rho)_{min} = -12$.
 
  For the grid results the volume weighted PDF is constructed simply by binning the grid cells according to the value of $\ln \rho$. To obtain a volume weighted PDF in SPH it is necessary to weight the contribution of each particle by the volume element associated with that particle, $m/\rho$, which for equal mass particles is simply inversely proportional to the density. To construct the PDF we therefore bin each particle according to the value of $\ln \rho$ and add a contribution of $1/\rho$ to the bin, normalising the resultant PDF such that the integral over all bins (i.e, the total probability) is unity. This is different to the procedure used to construct density PDFs from SPH particles used both in the Potsdam comparison \citep{kitsionasetal09} (see however Fig.~11 in \citealt{kitsionasetal09}) and by previous authors \citep[e.g.][]{vsbpk03,klessen00,maclowetal98}, whereby the SPH results were first interpolated to a grid and a PDF constructed as above for the grid-based results. The main disadvantage to interpolating to the grid is that the part of the high density tail of the SPH calculation that falls below the grid scale is removed. To retain this tail requires that PDFs \emph{should not} be constructed from SPH particles by interpolating to a grid, though this is a perfectly valid procedure for computing volumetric quantities such as power spectra (see \S\ref{sec:pspec}).

 The PDFs thus constructed (Fig.~\ref{fig:pdflog}) show clearly a log-normal distribution in agreement with many previous calculations and with theoretical expectations (see above). Whilst the results are broadly similar for all calculations in the central regions around the mean density and in overall shape, clear differences may also be observed between codes and with resolution, particularly in the tails of the distribution (Fig.~\ref{fig:pdftails}). At the low density end (Fig.~\ref{fig:pdftails}) the grid results tend to show a wide, low density tail, with probability densities \emph{decreasing} with resolution. By contrast, the SPH results show a narrower low density tail with probability densities that \emph{increase} with resolution.
 
  Whilst both codes appear to be converging towards each other, it is clear that neither is well converged at the low density end ($\rho/\rho_{0} \lesssim 0.01$) at least for the resolutions used in this paper. Thus the low density tail should not be used to fit the PDF width from either grid or SPH codes alone at these resolutions. Instead we measure the PDF width using the best fit around the mean value (i.e., based on Fig.~\ref{fig:pdflin}). The resulting best fit log-normal distributions are plotted as dashed lines in Fig.~\ref{fig:pdflin}, corresponding to $b=0.33$ for the $512^{3}$ SPH results and $b=0.5$ for the $512^{3}$ grid results. As it turns out, the best fitting log-normal distributions also provide a good fit to the low density tails (Figures~\ref{fig:pdflog} and \ref{fig:pdftails}), though the same is not true at very high density (see Fig.~\ref{fig:pdftails}). 

  At the high density end the trend with resolution for both codes is for the PDF in the high density tail to increase slightly, increasing the PDF width. The \textsc{flash} results show a stronger trend with resolution at high densities and appear to converge towards the \textsc{phantom} results, with rough equivalence between the $512^{3}$ grid based results and the $128^{3}$ SPH results in resolving the high density tail, similar to what is observed for the evolution of maximum density in Fig.~\ref{fig:rhomax}. Interestingly, the SPH results appear converged between the $256^{3}$ and $512^{3}$ calculations for densities up to around $\rho/\rho_{0} \sim 10^{3}$ (though not to the best fit log-normal, see below), with the primary effect of the additional resolution at $512^{3}$ being to extend the tail to lower probabilities (but with similar overall width).

That the SPH results appear close to converged between the $256^{3}$ and $512^{3}$ and that the grid results are converging towards them suggests that a value of $b\approx 0.35-0.4$ would be the converged value. This is similar to the results of \citet{beetzetal08} for solenoidal forcing ($b=0.37$), and close to the prediction of $b=1/3$ for solenoidal forcing from the heuristic model for the $b$ parameter presented in \citet{FederrathKlessenSchmidt2008} and \citet{FederrathDuvalKlessenSchmidtMacLow2009}.

The reason for the discrepancy at high density merits some consideration given the importance of this regime in relation to star formation. The most straightforward conclusion is that the convergence we find is merely incidental and that performing calculations at higher resolution would resolve the remaining discrepancy. Certainly, resolution requirements on the PDF become greater at higher Mach numbers because of the stronger time-variability in the tails of the distribution. This is evident from the strong fluctuations in $\rho_{max}$ (Fig.~\ref{fig:rhomax}) that are up to an order of magnitude larger than the fluctuations shown in Figure 2 of \citet{kritsuketal07}. There may also be differences due to the random forcing algorithm. In particular \citet{FederrathDuvalKlessenSchmidtMacLow2009}, using the same (solenoidal) forcing algorithm employed here, also find a small deviation from log-normality at the high density end, though smaller than we find since they employ $1024^{3}$ elements and a lower Mach number (Mach 6). They also found strong deviations from log-normality when a compressible forcing was applied, indicating that the PDF at high densities is quite sensitive to the forcing employed. \citet{FederrathDuvalKlessenSchmidtMacLow2009} discuss intermittency as a cause of non-Gaussian PDF tails. They quantify and discuss the Mach number--density correlations as the key to non-Gaussian PDFs \citep[see][]{vs94,pvs98}. \citet{falgaroneetal94} and \citet{hilyblantetal08} also find strong intermittent fluctuations in their molecular cloud observations and attribute them to a fundamental property of turbulence, i.e., intermittency \citep[see, e.g.,][]{sl94}.

 In principle it is also possible to compute the PDFs from the tracer particle density field calculated with the SPH summation. However we find that the resulting PDFs show a strong deviation from a log-normal distribution, particularly in the high density tail (much stronger than those seen in Fig.~\ref{fig:pdftails} and in the opposite direction), due to the manner in which tracer particles tend to cluster in high density regions at later times (see discussion in \S\ref{sec:slices} and right panels in Fig.~\ref{fig:slicerho_t10}).

\subsection{Power spectra}
\label{sec:pspec}

\citet{pn02} derive a relationship between the mass distribution of dense cores to the slope of the kinetic energy power spectrum in supersonic, super-Alfv\'enic turbulence via the relation
\begin{equation}
N(m) d\log m \propto m^{-3/(4-\beta)} d\log m,
\end{equation}
where $\beta$ is the slope of the kinetic energy power spectrum assumed to be a power law of the form
\begin{equation}
E(k) \propto k^{-\beta},
\end{equation}
where $\beta$ would be $\sim 5/3$ according to the \citet{kolmogorov41} phenomenology for incompressible turbulence. Supersonic turbulence is generally found to have a power spectrum closer to the completely pressure-free shock-dominated turbulence produced by solving Burgers' equation, $\beta = 2$, implying dominance of the nonlinear advection term $(v\cdot \nabla) v$ over the pressure gradient in the equation of motion. A value consistent with the observed \citet{salpeter55} slope for the IMF of $N(m) d\log m \propto m^{-4/3} d\log m$ requires $\beta \approx 1.75$, which \citet{pn02} suggest arises from supersonic MHD turbulence in the super-Alfv\'enic regime.

 Recently \citet{kritsuketal07} have suggested that Kolmogorov scaling may be applicable for highly compressible turbulence by assuming that the mean volume energy transfer rate $\rho v^{2} v/l$, is constant, implying that $E(k) \propto k^{-5/3}$ holds for the variable $w \equiv (\rho^{1/3} {\bf v})$ rather than simply for the velocity. We thus consider power spectra of both quantities.

\begin{figure}
\begin{center}
 \includegraphics[width=\columnwidth]{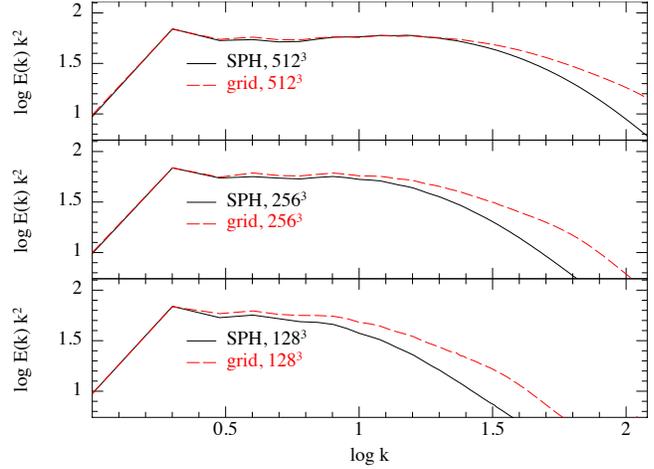}
 \caption{Velocity power spectra ($\frac12 v^{2}$), shown compensated by $k^{2}$ as an average over $81$ snapshots evenly spaced between $t/t_{d}=2$ and $t/t_{d}=10$ for the SPH (solid, black) and grid (dashed, red) calculations at the three different resolutions, as indicated. For calculations at or above $256^{3}$ computational elements in either code the results in the scaling range  $7\lesssim k \lesssim 11$ are consistent with a slope slightly shallower than Burgers' value of $k^{-2}$, between $k^{-1.93}$ and $k^{-1.98}$ depending on whether or not the $7\lesssim k \lesssim 11$ region is interpreted as inertial range or a bottleneck effect.}
\label{fig:ekin}
\end{center}
\end{figure}

\begin{figure}
\begin{center}
 \includegraphics[width=\columnwidth]{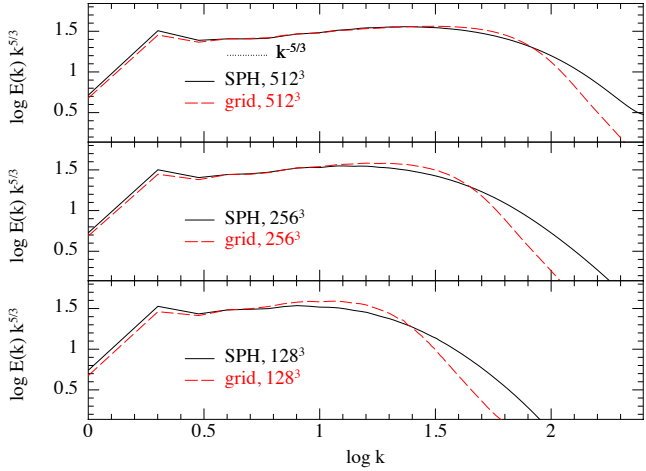}
 \caption{Density-weighted velocity power spectra, as in Fig.~\ref{fig:ekin} but for the quantity $(\rho^{1/3}{\bf v})$ instead of the velocity field. There is tentative evidence of a flat portion in the compensated spectra at resolutions above $256^{3}$ in either code, suggesting a small scaling range in the region of $4 \lesssim k \lesssim 6$ (dotted line) consistent with a Kolmogorov-like $k^{-5/3}$ scaling for this quantity as suggested by \citet{kritsuketal07}. This also suggests that $512^{3}$ particles/grid cells is the minimum resolution requirement to determine the power spectrum slope in the inertial range in either code \citep[see also][Fig.~C.1]{FederrathDuvalKlessenSchmidtMacLow2009}.}
\label{fig:rho3}
\end{center}
\end{figure}

\begin{figure}
\begin{center}
 \includegraphics[width=\columnwidth]{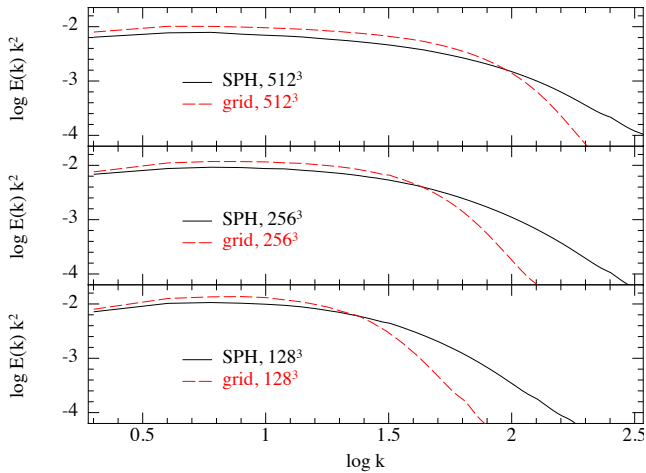}
 \caption{Power spectra, as in Figs.~\ref{fig:ekin} and \ref{fig:rho3}, but here for the density field, computed as the spectrum of the density fluctuations $\delta\rho\equiv \rho - \bar{\rho}$. Clearly the SPH code provides increased resolution in the density field at high $k$ compared to the grid-based code for the same number of computational elements. The spectra have been compensated by $k^{2}$, though it is evident that the slope of the density spectrum is not well described by a single power-law.}
\label{fig:pspecrho}
\end{center}
\end{figure}

 We have computed the power spectra for each code directly from gridded data using the same analysis script for both codes. For the SPH code, this means that the results have first been interpolated to grids of size $256^{3}$, $512^{3}$ and $512^{3}$ cells for the $128^{3}$, $256^{3}$, $512^{3}$ particle calculations respectively. The interpolation has been performed using a routine from the \textsc{splash} visualisation code, the details of which are described in \citet{splashpaper}. The main disadvantage of interpolating SPH data to a grid is that any resolution in the SPH code on scales smaller than the grid scale is lost. This is most obvious when comparing quantities such as the maximum density on the particle data compared to the maximum density on the interpolated grid. For example the maximum density interpolated onto a $512^{3}$ grid ($\max (\rho/\rho_{0}) \sim 3-4 \times 10^{2}$) is a factor of $\sim 3-4$ lower than the maximum density from the particles ($\max (\rho/\rho_{0}) \sim 1-2 \times 10^{3}$) for snapshots from the $512^{3}$ SPH calculation, which would remove some of the information from the high density tail of the PDF as discussed above. Whether or not the interpolation procedure affects the power spectrum calculation can be determined simply by comparing the results from interpolations to different sized grids. We find that for the power spectrum, as one might expect given that it is a volumetric measurement, the power spectra are identical for different grid sizes apart from $k$'s very close to the grid scale.

 Power spectra
\begin{equation}
E({\bf k}) = \frac12 \tilde{\bf w}(\bf k)^{2},
\label{eq:ekinft}
\end{equation}
for an arbitrary vector field ${\bf w}$ are constructed from the 3D Fourier transform,
\begin{equation}
\tilde{\bf w}({\bf k}) = \frac{1}{(2\pi)^{3}} \int_{V} {\bf w}({\bf x}) e^{-i2\pi {\bf k}\cdot{\bf x}} {\rm d}^{3}{\bf x},
\label{eq:ft}
\end{equation}
The angle-averaged power spectrum was then obtained by the standard procedure of summing E({\bf k}) in bins according to the norm of the wavenumber, $\vert k \vert$ \citep[as in e.g.][]{kitsionasetal09}.

\subsubsection{Volume weighted velocity power spectra}
 The velocity power spectrum (that is, where ${\bf w}\equiv{\bf v}$), averaged over 81 snapshots as described above for the PDFs, is shown --- compensated by $k^{2}$ --- in Fig.~\ref{fig:ekin} for the three SPH calculations (solid black lines) and for the three grid calculations (dashed red lines). The results, similar to previous studies, show a peak at the driving scale $k\sim 2$, a power law slope (flat in these compensated spectra) between $k\sim 4$ and up to $k\sim 12$ in the highest resolution calculations and an extended dissipative tail at large $k$. Both codes show a power law slope close to the pressure-free Burgers model slope of $k^{-2}$. The trend with resolution is for both codes to tend towards a slightly shallower slope, with the highest resolution calculations consistent with $\sim k^{-1.95}-k^{-1.98}$ for both codes at $512^{3}$, depending on whether or not the results for $7 < k < 12$ are interpreted as inertial range or as a bottleneck effect. The extent of the power-law scaling range of the velocity power spectrum appears to be very similar in both codes at equivalent (number of grid cells = number of particles) resolutions, though the SPH results show a faster drop-off towards higher $k$ in the dissipative tail in agreement with the results by \citet{kitsionasetal09}.
 
  These results are entirely consistent with the power law slopes obtained in Figure 2 of \citet{padoanetal07}. However they differ strongly from the low resolution SPH and TVD results shown in Fig.~2 of \citet{bpetal06}, shown compensated in Fig.~8 of \citet{padoanetal07}. In particular our SPH results show power law slopes consistent with $\beta=2$ or \emph{shallower}, which is in stark contrast to the $\beta=2.7-2.9$ obtained by \citet{bpetal06}, albeit at lower Mach numbers (6 and 3 respectively). This can be understood primarily as an effect of the numerical resolution, since we observe a significantly steeper slope in our low resolution SPH calculations shown in Fig.~\ref{fig:ekin}. As noted in the introduction however, the resolution in our `low' resolution SPH calculation, employing 2.1 million particles, is already an order of magnitude higher than than the $\sim 200,000$ particles used for most of the calculations in \citet{bpetal06} (they perform one `high' resolution calculation using $\approx 10$ million particles, the power spectrum of which is not shown in their paper, though it is used to derive a core mass distribution). We also find excellent agreement between both the SPH and grid-based results which is in contrast to \citet{bpetal06} where the measured slopes differ between their codes which had been attributed to the very different properties of the SPH and TVD schemes used in their paper.

\subsubsection{Density weighted velocity power spectra ($\rho^{1/3}{\bf v}$)}
 Fig.~\ref{fig:rho3} shows the time-averaged power spectrum, computed as above, of the quantity ${\bf w}\equiv \rho^{1/3}{\bf v}$, which according to the hypothesis of \citet{kritsuketal07} is the quantity that for supersonic turbulence should show a Kolmogorov-like scaling of $k^{-5/3}$. We have therefore compensated the spectra by $k^{5/3}$ in order to assess whether or not this can be supported on the basis of our calculations. The spectra in Fig.~\ref{fig:rho3} show similar generic features to those observed in Fig.~\ref{fig:ekin}.
 
 
 The lower resolution calculations ($128^{3}$ and $256^{3}$) in both codes show slopes that appear shallower than $\beta = 5/3$, though the convergence of both codes is towards a steeper slope with resolution. In particular the $512^{3}$ calculations with both codes show a small, flat region in the compensated spectrum between $k\sim 4$ and $k\sim 6$ that may be interpreted as a resolved inertial range (shown by the dotted black line) consistent with the \citet{kritsuketal07} scaling, with a `bottleneck effect' for $k > 6$ extending into the dissipative tail. For the grid code, this is consistent with the findings of \citet{kritsuketal07} and \citet{FederrathDuvalKlessenSchmidtMacLow2009} that $512^{3}$ is roughly the minimum resolution required to resolve the inertial range, and we conclude that a similar requirement holds for SPH.
 
  Interestingly, the difference between codes in rate of drop off in the high-$k$ dissipative tail for the $\rho^{1/3}{\bf v}$ spectrum is the reverse of what occurs for the velocity field alone (comparing to Fig.~\ref{fig:ekin}). That is, for $\rho^{1/3}{\bf v}$ the grid-based results drop off much faster at high-$k$ than the SPH code, whereas the SPH code drops off faster in the velocity spectrum. We interpret this as being due to the fact that SPH has better \emph{mass} resolution for equal numbers of computational elements (reflected in the density field) but worse \emph{volume} resolution (reflected in the velocity field). This is further evident in the spectrum of the density field alone, discussed below.

\subsubsection{Density power spectra}
 Fig.~\ref{fig:pspecrho} shows the power spectra of the density field, computed as the spectrum of density fluctuations $\delta \rho \equiv \rho - \bar{\rho}$ such that the integral under the power spectrum gives the density variance. Whilst not well fit by a single power law we show the spectra compensated by $k^{2}$ to facilitate the comparison. The overall shape of the spectra is similar between codes on large scales (small $k$), though it is clear that the SPH code shows a much higher resolution in the density field, falling much more slowly at high $k$ compared to the grid results at comparable resolutions and explaining the results in Fig.~\ref{fig:rho3} as intermediate between Figures~\ref{fig:ekin} and \ref{fig:pspecrho}. The SPH density spectrum at $128^{3}$ particles appears better resolved than the $256^{3}$ grid spectrum though not as well resolved as the $512^{3}$ grid spectrum, lying somewhere between the two.
 
\begin{figure*}
 \includegraphics[width=\columnwidth]{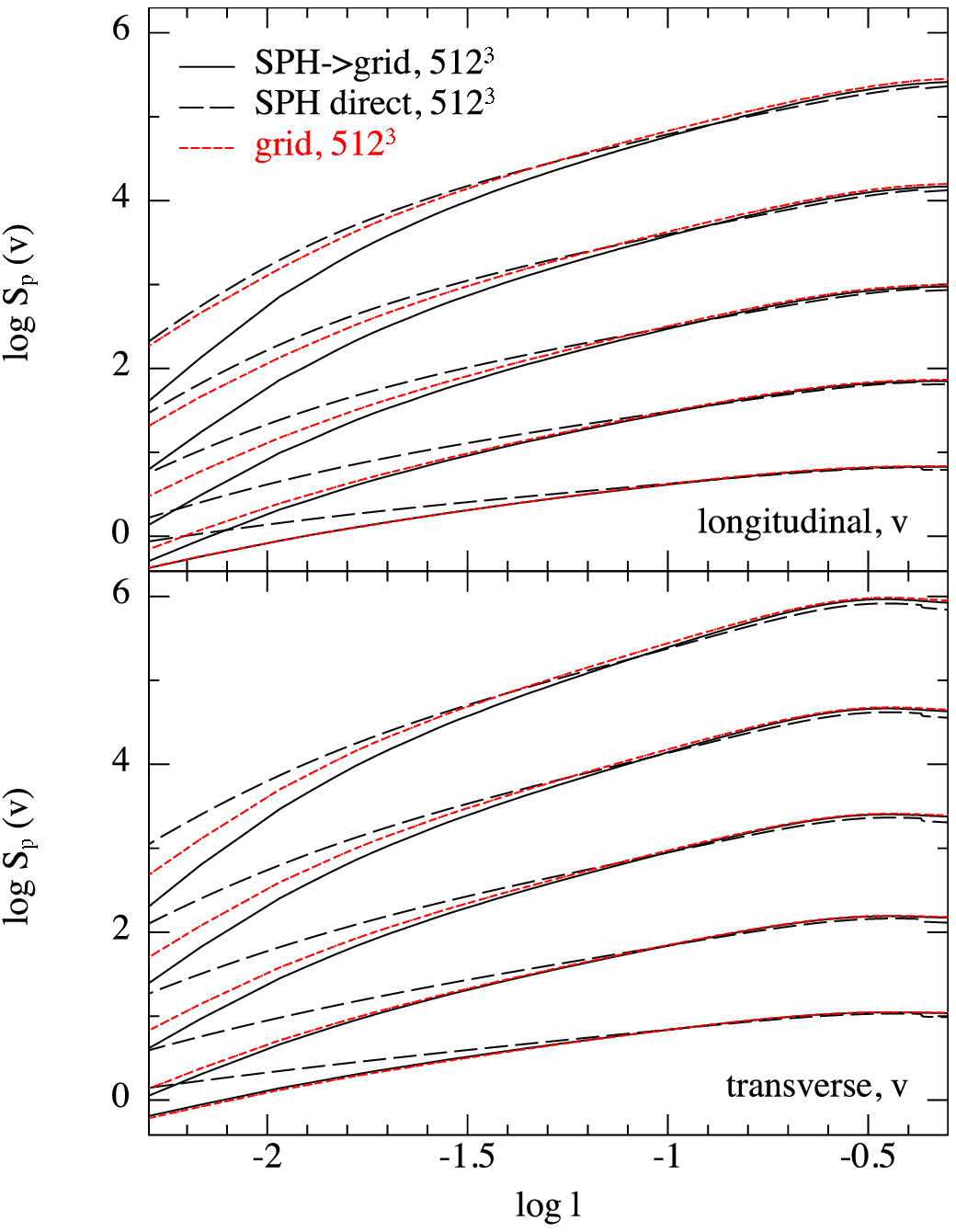}
 \hspace{0.05\columnwidth}
 \includegraphics[width=\columnwidth]{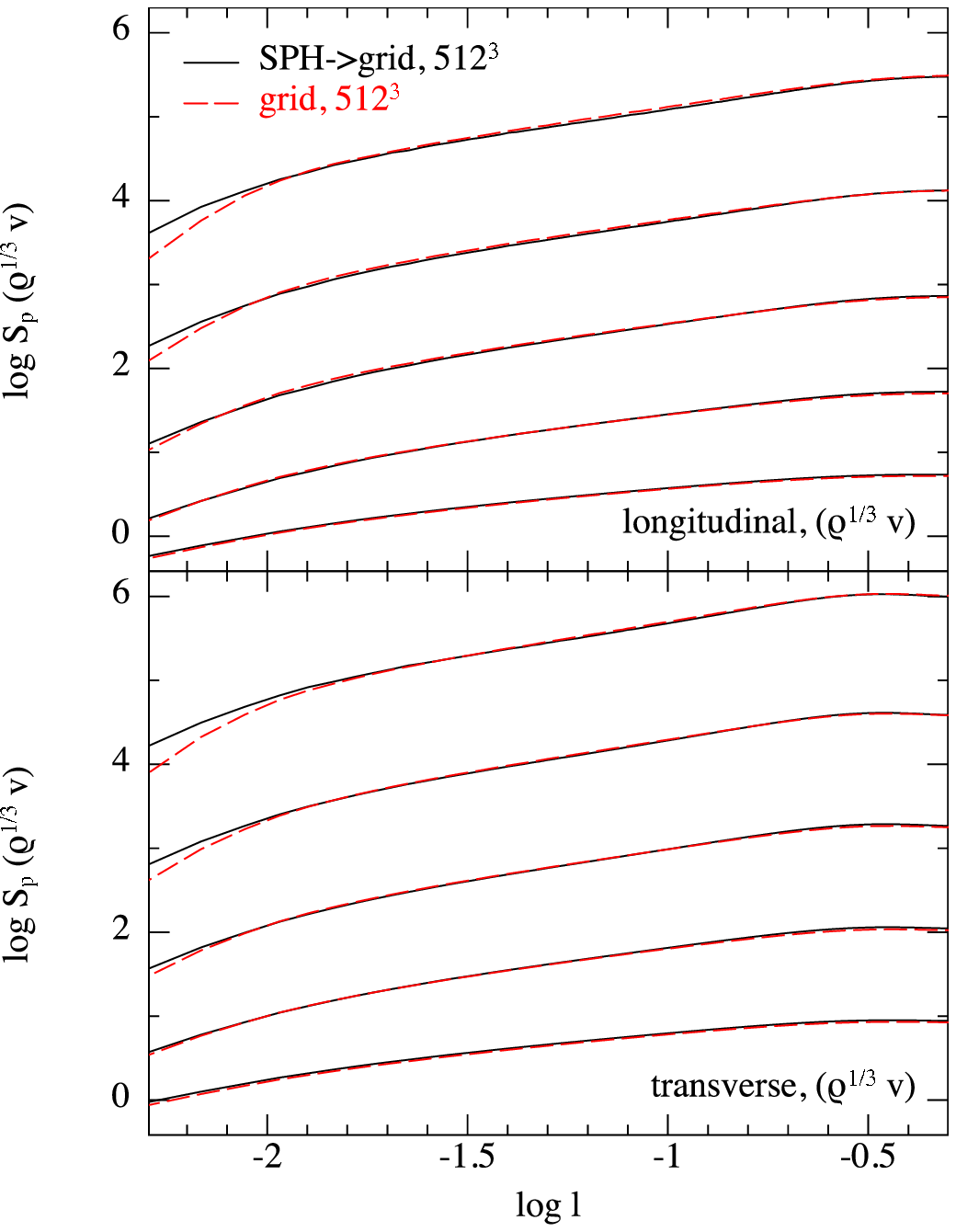}
 \caption{Structure functions $S_{p} \equiv \langle\vert {\bf w}({\bf r} + l) - {\bf w}({\bf r}) \vert^{p} \rangle$ as a function of spatial separation (the lag, $l$), for orders $p=1$,$2$,$3$,$4$ and $5$, showing the time average over $81$ snapshots evenly spaced from $t_{d} = 2$ to $t_{d} = 10$ in each case. Structure functions are shown for the longitudinal (${\bf w}\parallel {\bf l}$, top) and transverse (${\bf w}\perp {\bf l}$, bottom) components of the velocity field, ${\bf w} \equiv {\bf v}$ (left) and for the mass-weighted velocity ${\bf w} \equiv \rho^{1/3} {\bf v}$ (right). The results from the \textsc{flash} (grid) $512^{3}$ calculation are given by the dashed red lines whilst the \textsc{phantom} (SPH) results at $512^{3}$ are shown by the solid black lines, where these have first been interpolated to a $512^{3}$ grid in order to be analysed identically to the grid results. Results for the velocity field computed directly from the SPH particles for the $512^{3}$ calculation are plotted on the left figure using dashed black lines.}
\label{fig:sfall}
\end{figure*}

\begin{figure*}
 \includegraphics[width=\columnwidth]{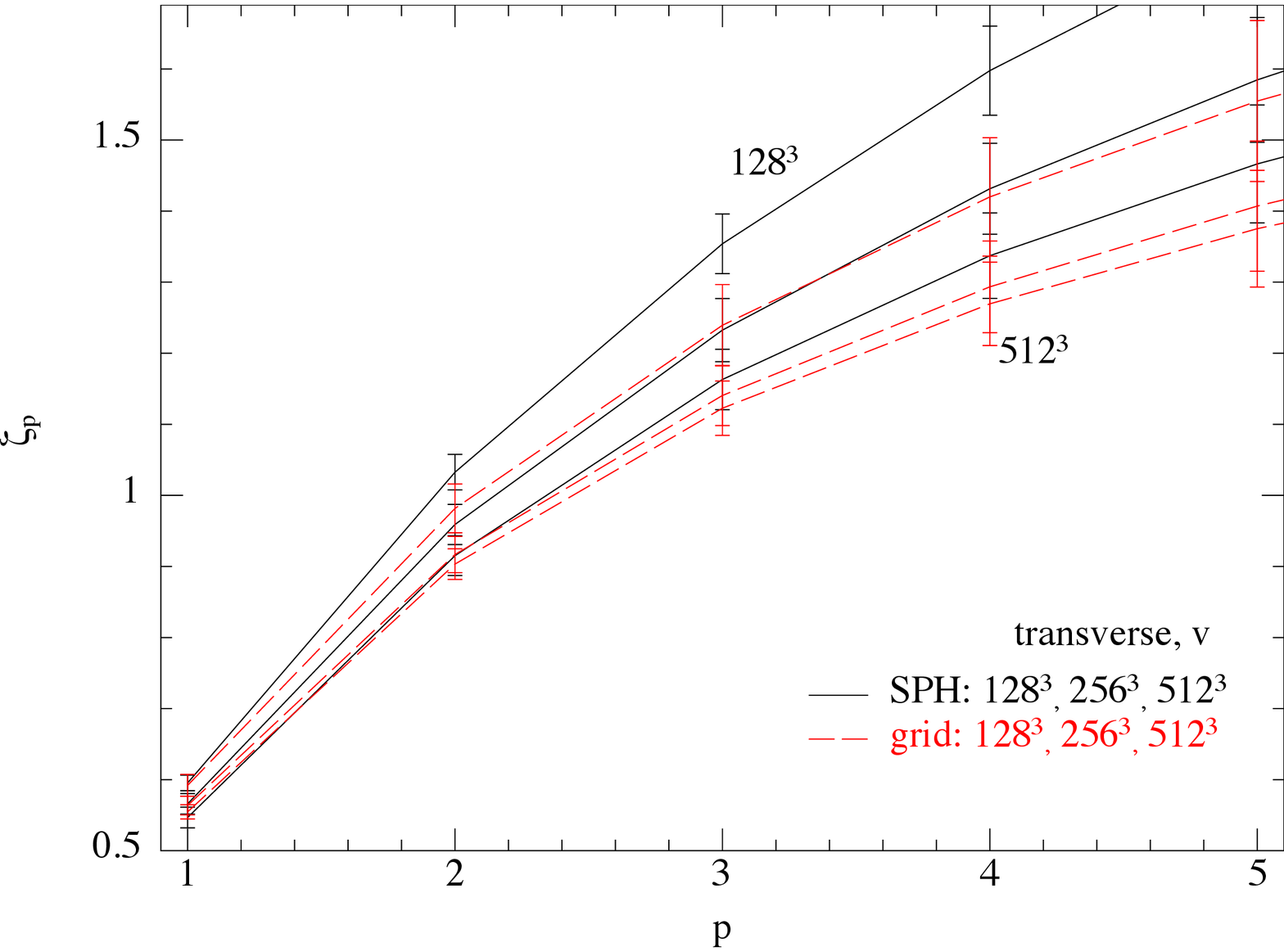}
 \hspace{0.05\columnwidth}
 \includegraphics[width=\columnwidth]{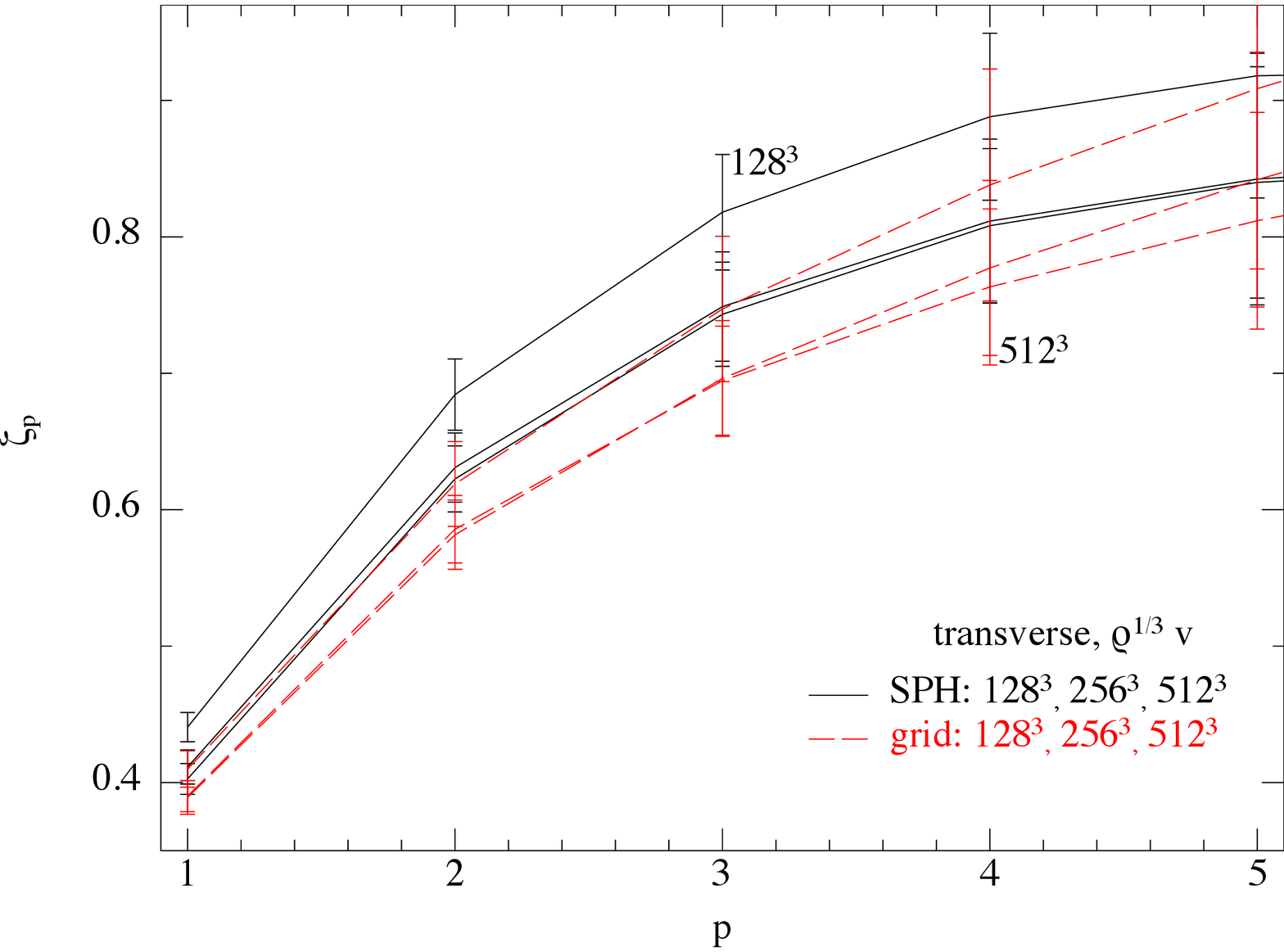}
 \caption{Time-average of the measured structure function slope $\zeta_{p}$, where $S_{p} \propto l^{\zeta_{p}}$ for the transverse velocity (left) and $\rho^{1/3} {\bf v}$ (right), averaged from $81$ snapshots evenly spaced between $t/t_{d}=2$ and $t/t_{d}=10$, and plotted as a function of the structure function order $p$. The solid black lines correspond to the \textsc{phantom} (SPH) calculations whilst the red dashed lines correspond to the \textsc{flash} (grid) calculations, each shown at 3 different resolutions (the slopes decrease with resolution in both codes). Error bars show the $1\sigma$ errors in the time-average of the measured slope. The corresponding time average of the structure functions themselves are shown in the lower panels of Fig.~\ref{fig:sfall}.}
\label{fig:sfscalev}
\end{figure*}



\subsection{Structure Functions}
 We have measured the structure functions,
\begin{equation}
S_{p}(l) \equiv \langle\vert {\bf w}({\bf r} + l) - {\bf w}({\bf r}) \vert^{p} \rangle
\label{eq:sfunc}
\end{equation}
from our calculations, where in this paper we show orders up to 5, i.e., $p=1,2,3,4$ and $5$ and the variable ${\bf w}$ is either the velocity ${\bf w}\equiv {\bf v}$ or the quantity ${\bf w}\equiv (\rho^{1/3}{\bf v})$ as for the computation of the power spectra in the previous section. We compute intrinsically three dimensional structure functions (rather than the approximate `directionally split' versions used e.g. in the KITP comparison) by sampling pairs of either grid cells or SPH particles, computing the relative velocity difference and adding the result (the absolute value to the power $p$) to the appropriate bin corresponding to the spatial separation of the points (the ``lag'' --- given here in computational units i.e. in terms of the box length L). Dividing by the total number of contributions to each bin produces the average, as indicated by the angle brackets $\langle ... \rangle$. We have computed the structure functions for both the transverse ({\bf w} perpendicular to the line ${\bf l}$ joining the pair of points) and longitudinal ({\bf w} parallel to ${\bf l}$) components of ${\bf w}$ in each case.

 Calculation of the structure functions is computationally expensive, in principle requiring a summation over $\mathcal{O}(N^{2})$ pairs of points. A more efficient calculation can be made by selecting only a subsample of points. We achieve this by randomly selecting a fixed number of points (either grid cells or particles) and computing the pair-wise interaction of these sample points with all points in the computational domain, with a constraint to achieve similar sampling of all lags. For the structure function calculations we typically use $\sim 250,000$ sample points. For the SPH results we have firstly computed structure functions by interpolating the SPH data to a grid, as described above for the power spectra. Whilst this ensures that the analysis step is identical to the grid-based results, this procedure has the disadvantage that the SPH densities and velocities are smoothed by the interpolation step. For comparison we have therefore computed structure functions also directly from the particle data. In the particle version we iterate the structure functions to completion by progressively increasing the number of sample points by a factor of 2 and check that the error between the final result and the structure functions using half the number of sample points is small (typically we require the RMS error on the highest order structure function to be $< 1\%$). In practise this produces a similar sample size to that used for the grids i.e., a few hundred thousand sample points for a $512^{3}$ calculation, but can be more efficient if the velocity field is well ordered (e.g. at early times).

\subsubsection{Velocity structure functions}
 The structure functions for the velocity field (${\bf w}\equiv {\bf v}$) at all computed orders ($p=1$..$5$) are shown in the left hand side of Fig.~\ref{fig:sfall} for the  $512^{3}$ calculations with both codes. The structure functions show the same general features in both codes, showing slopes that appear power-law like in the range $l \sim 0.03$ to $l \sim 0.3$ (though see below), a flat structure around the driving range ($0.3 \lesssim l \lesssim 0.5$) and a fall-off in the dissipative tail ($l \lesssim 0.03$). The transverse structure functions appear flatter around the driving scale compared to the longitudinal version, consistent with the fact that we have used purely solenoidal forcing. Differences between codes mainly appear in the dissipative tail, as observed also in the power spectra (Figures~\ref{fig:ekin} and \ref{fig:rho3}). In the velocity structure functions, as with the velocity power spectrum, it appears that the SPH results fall off faster in the dissipative tail (comparing solid and dot-dashed lines in the left hand part of Fig.~\ref{fig:sfall}), though the results computed directly from the particles (dashed lines) appear to show a slower fall-off, indicating that at least some of this may be an artefact of the interpolation procedure. However the structure functions computed directly from the particles also show much shallower slopes at the lowest orders ($p=1$..$4$), so it is not clear that the structure functions computed in this manner are truly comparable.
 

\subsubsection{Density weighted velocity structure functions ($\rho^{1/3}{\bf v}$)}
 Following \citet{kritsuketal07b}, we have also computed structure functions for ${\bf w} \equiv (\rho^{1/3} {\bf v})$. These are shown on the right hand side of Fig.~\ref{fig:sfall}, showing --- given the above --- only the versions computed identically from a grid (i.e., interpolated in the SPH case). Again, the qualitative features are in good agreement between the two codes --- in particular the structure functions for $\rho^{1/3}{\bf v}$ show a power-law scaling range over a much wider range of distances, $0.01 \lesssim l \lesssim 0.3$. As was the case for the power spectra (see Fig.~\ref{fig:rho3}), with the density-weighting the grid code (solid lines) shows a faster fall-off in the dissipative tail compared to the SPH code (dashed lines), in contrast to the pure velocity version (comparing the right hand and left hand panels in Fig.~\ref{fig:sfall}).
 
\subsection{Structure function scaling}
 A more quantitative analysis can be made by performing a least squares fit to determine the structure function slope $\zeta_{p}$ assuming $S_{p}\propto l^{\zeta_{p}}$ for each order $p$. The results of such a fit to the time-averaged structure functions shown in Fig.~\ref{fig:sfall} --- performed over a fixed $\Delta l$ in the range $0.1 < l < 0.15$ (see below) --- are shown in Fig.~\ref{fig:sfscalev}. Plotted are the fitted slope $\zeta_{p}$ at each order as a function of $p$, for both the velocity structure functions (left panel) and the density-weighted versions (right panel). The results from the SPH code, computed from the grid-interpolated structure functions, are shown in black, whilst the {\sc flash} (grid) results are shown in red. In order to quantify the time variability in the value of the slope, we have also computed the standard deviation in the structure function slope at each order computed for each of the 81 snapshots discussed above. The $1-\sigma$ deviations are plotted as the error bars in Fig.~\ref{fig:sfscalev}.

 For the velocity structure functions (left panel of Fig.~\ref{fig:sfscalev}), reasonable agreement between codes is seemingly obtained for the two lowest order structure functions ($p=1$ and $2$) at the highest resolution employed ($512^{3}$, though see below). The grid code (red, dashed lines) appears close to (though not fully) self-converged at these orders (but not for $p > 2$) whilst the SPH code (solid, black lines) appears only close to convergence at the lowest order. The situation appears slightly better for the $\rho^{1/3}$-weighted structure functions (right panel of Fig.~\ref{fig:sfscalev}), where the grid code (red, dashed lines) shows apparent convergence in the slopes up to $p=3$ between the $256^{3}$ and $512^{3}$ calculations. The SPH results (black, solid lines) appear to show only a small change between the measured slopes at all orders between the $256^{3}$ and $512^{3}$ results, though given the remaining disagreement between codes even at $512^{3}$ it is not clear that true convergence has been reached in either.

 The somewhat large caveat to the results shown in Fig.~\ref{fig:sfscalev} is our finding that the determined slope depends strongly on the range of scales ($\Delta l \equiv l_{max} - l_{min}$) used to perform the least squares fit. This is discussed in more detail in Appendix~\ref{sec:sfslopefit} and demonstrated by Fig.~\ref{fig:slopefitting} that shows the dependence of $\zeta_{p}$ on the employed $\Delta l$ for the transverse structure functions. It is clear from the fact that the measured slope steepens as $\Delta l$ is increased that relatively little true power-law scaling range is apparent in either code at the resolutions employed. This is particularly true for the velocity structure functions (left panel of Fig.~\ref{fig:slopefitting}) where either a very short ($\Delta l \lesssim 0.02$) scaling range exists or none at all. The situation is marginally improved for the density-weighted structure functions (right panel of Fig.~\ref{fig:slopefitting}), where the measured slope is less dependent on the fitting range (also apparent from Fig.~\ref{fig:sfscalev}). However there remains a significant dependence on $\Delta l$ even for the lowest order $p=1$.  Furthermore, it is clear that any nominal convergence between $256^{3}$ and $512^{3}$ results in individual codes seen in Fig.~\ref{fig:sfscalev} \emph{only} occurs if a relatively short range in $l$ ($\Delta l \lesssim 0.05$, i.e. $0.1 \lesssim l < 0.15$) is used to perform the fit (and, as noted above, the measured slopes in this range also do not agree between codes). It is therefore difficult to assert that either code produces a resolved scaling in either the velocity or density weighted structure functions at \emph{any order} at these resolutions.

 Given the above, we have not attempted a detailed comparison with phenomenological models for the structure function scaling, such as those proposed by \citet{sl94} (revised for supersonic turbulence by \citealt{boldyrev02} and \citealt{sfk08}). Nevertheless, although the true structure function scaling remains to be determined, it is clear that the scaling for both $v$ and $\rho^{1/3} v$ does not match either the \citet{kolmogorov41} ($\zeta_{p} \propto p/3$) or Burgers ($\zeta_{p} = 1$ for $p > 1$) expectations, and also does not match the \citet{boldyrev02} theory, though the relative scaling we have measured does compare reasonably well to the parameters suggested by \citet{sfk08}.

\section{Discussion and Conclusions}
\label{sec:discussion}

 In this paper we have performed a detailed comparison of the statistics of supersonic turbulence using high resolution numerical simulations with two fundamentally different codes: \textsc{flash}, an Eulerian grid-based code and \textsc{phantom}, a Lagrangian particle-based SPH code. Despite the very different numerical methods we find in general very good agreement between the codes on the many aspects of supersonic turbulence, though it is clear that neither code shows results that are fully converged at $512^{3}$ except for a small scaling range in the velocity power spectrum with both codes and some indication of convergence near the peak and at the high density end of the PDF in the SPH case.
 
  We find good agreement in the fact that hydrodynamic turbulence at Mach$~10$ has a velocity power spectrum with a slope around $E(k) \propto k^{-1.95}$, very close to Burger's value of $E(k) \propto k^{-2}$, confirming many previous results using only grid-based codes \citep[e.g.][]{padoanetal07,FederrathDuvalKlessenSchmidtMacLow2009}. At the highest resolution employed, both codes support the idea that the power spectrum of the mixed quantity $\rho^{1/3}{\bf v}$ shows a Kolmogorov-like scaling in the power spectrum of $k^{5/3}$, as proposed by \citet{kritsuketal07}. Both codes agree that the PDF of the logarithm of the density shows a distribution that is log-normal to good approximation for densities around $3-4$ orders of magnitude either side of the mean density, with the factor $b$ relating the width to the Mach number measured to lie between $b=0.33$ to $b=0.5$, with the converged value expected to lie somewhere around $b\approx 0.4$. However, both codes also show that there are significant deviations from a log-normal distribution at high density, as also found by \citet{FederrathDuvalKlessenSchmidtMacLow2009}.
  
Our conclusions regarding resolution and computational requirements are as follows: 
\begin{enumerate}
\item For measuring volumetric statistics such as the power spectrum slope and structure function scaling, we find that SPH and grid codes give roughly comparable results when the number of SPH particles is approximately equal to the number of grid cells. This means that SPH codes are not well suited to measuring such quantities since for this kind of problem an SPH code will be significantly more expensive than a uniform-grid implementation at similar numbers of computational elements (we find the cost of the SPH calculations with \textsc{phantom} similar to the cost of running the grid code (\textsc{flash}) at twice the resolution, i.e., with $256^{3}$ particles $\approx 512^{3}$ grid cells in terms of CPU time).
\item On the other hand the SPH code was found to be better at resolving dense structures, giving maximum densities at a resolution of $128^{3}$ particles that were similar to the maximum densities resolved in the grid code at $512^{3}$ cells, reflected also in the high density tail of the PDF. SPH is therefore a more efficient method in this regard, which is an important reason why it is frequently used for studying star formation (for which a grid based code requires adaptive mesh refinement, adding a significant computational overhead).
\item At comparable resolutions ($N_{part} \approx N_{cells}$), SPH (\textsc{phantom}) appears to be more dissipative (steeper dissipative tails in the power spectrum and structure functions) in the velocity field, but the reverse is true for the statistics of the density-weighted quantity $\rho^{1/3} {\bf v}$ where the grid (\textsc{flash}) results appear more dissipative. We attribute the former to the greater sophistication of the shock capturing scheme in the grid-based code, and the latter to the better resolved power in the density field provided by the SPH code.
\item The absolute values of the structure function slopes are not converged in either grid or SPH at $512^{3}$, requiring much higher resolution in order to make a meaningful comparison with scaling models.
\item In order to accurately simulate supersonic turbulence in SPH it is important to ensure that sufficient $\beta$-viscosity is applied to prevent particle interpenetration in shocks. At Mach~10 we require $\beta_{visc} = 4$ instead of the usual $\beta_{visc} = 2$.
 \item We find that calculation of the sub-grid density field from the tracer particle distribution using an SPH summation can significantly enhance the resolution of high density structures from the grid-based results. However it is unclear whether or not the statistics of these sub-grid structures can be used in a meaningful way, because of the manner in which tracer particles tend to cluster in high density regions.
\end{enumerate}

 The above conclusions mean that the differences between the SPH and grid based results discussed by \citet{padoanetal07}, at least in terms of the power spectrum slope, can be understood as a consequence of the low resolution employed in the SPH calculations rather than being due to an intrinsic difficulty with the method. Given this, it is also likely that the conclusions drawn by \citet{bpetal06} regarding the presence or otherwise of an emergent power law slope in the core mass distribution should also be treated with caution. As a follow-up to this paper it would be interesting to examine the properties of dense clumps (or `cores') in our calculations using a clump-finding algorithm in a similar manner to \citet{bpetal06} and \citet{padoanetal07} to see whether or not these differences can also be reconciled by improved resolution in the SPH results.


\section*{Acknowledgements}
We thank the anonymous referee for comments that have substantially improved the paper and Aake Nordlund for kind provision of some of the analysis routines used to make a preliminary comparison. This work was initiated whilst CF was visiting Monash University under the Go8/DAAD Joint Research Co-operation Scheme grant led by Rosemary Mardling and Rainer Spurzem. CF is grateful for support from Ralf Klessen during the initial steps of this project. CF acknowledges financial support from the International Max Planck Research School for Astronomy and Cosmic Physics (IMPRS-A) and the Heidelberg Graduate School of Fundamental Physics (HGSFP: funded by the Excellence Initiative of the German Research Foundation DFG GSC 129/1). Figures were produced with {\sc splash} \citep{splashpaper}, using the new giza/cairo backend written by James Wetter. Calculations using \textsc{phantom} were run using the NCI supercomputing facilities and on a dedicated node of the Monash Sun Grid, ably managed by Philip Chan and the e-research centre at Monash without whom this research would not have been possible. The \textsc{flash} simulations used computational resources from the HLRB-II project grant pr32lo at the Leibniz Rechenzentrum Garching. The software used in this work was in part developed by the DOE-supported ASC / Alliance Center for Astrophysical Thermonuclear Flashes at the University of Chicago.

\bibliography{sph,starformation,turbulence,chfeder}

\begin{thebibliography}{}

\bibitem[\protect\citeauthoryear{{Ballesteros-Paredes}, {Gazol}, {Kim},
  {Klessen}, {Jappsen} \& {Tejero}}{{Ballesteros-Paredes}
  et~al.}{2006}]{bpetal06}
{Ballesteros-Paredes} J.,  {Gazol} A.,  {Kim} J.,  {Klessen} R.~S.,  {Jappsen}
  A.-K.,    {Tejero} E.,  2006, ApJ, 637, 384

\bibitem[\protect\citeauthoryear{{Ballesteros-Paredes}, {Klessen}, {Mac Low} \&
  {Vazquez-Semadeni}}{{Ballesteros-Paredes} et~al.}{2007}]{bpetal07}
{Ballesteros-Paredes} J.,  {Klessen} R.~S.,  {Mac Low} M.-M.,
  {Vazquez-Semadeni} E.,  2007, in {Reipurth} B.,  {Jewitt} D.,   {Keil} K.,
  eds, Protostars and Planets V, Univ. Arizona Press, Tucson, AZ pp 63--80

\bibitem[\protect\citeauthoryear{{Bartosch}}{{Bartosch}}{2001}]{bartosch01}
{Bartosch} L.,  2001, Int. J. Mod. Phys. C, 12, 851

\bibitem[\protect\citeauthoryear{{Beetz}, {Schwarz}, {Dreher} \&
  {Grauer}}{{Beetz} et~al.}{2008}]{beetzetal08}
{Beetz} C.,  {Schwarz} C.,  {Dreher} J.,    {Grauer} R.,  2008, Phys. Let. A,
  372, 3037

\bibitem[\protect\citeauthoryear{{Berger} \& {Colella}}{{Berger} \&
  {Colella}}{1989}]{BergerColella1989}
{Berger} M.~J.,  {Colella} P.,  1989, J. Comp. Phys., 82, 64

\bibitem[\protect\citeauthoryear{{Boldyrev}}{{Boldyrev}}{2002}]{boldyrev02}
{Boldyrev} S.,  2002, ApJ, 569, 841

\bibitem[\protect\citeauthoryear{{Brunt}}{{Brunt}}{2010}]{brunt10}
{Brunt} C.~M.,  2010, A\&A, 513, A67

\bibitem[\protect\citeauthoryear{{Brunt}, {Federrath} \& {Price}}{{Brunt}
  et~al.}{2010}]{bfp10a}
{Brunt} C.~M.,  {Federrath} C.,    {Price} D.~J.,  2010, MNRAS, 403, 1507

\bibitem[\protect\citeauthoryear{{Calder}, {Fryxell}, {Plewa}, {Rosner},
  {Dursi}, {Weirs}, {Dupont}, {Robey}, {Kane}, {Remington}, {Drake}, {Dimonte},
  {Zingale}, {Timmes}, {Olson}, {Ricker}, {MacNeice} \& {Tufo}}{{Calder}
  et~al.}{2002}]{CalderEtAl2002}
{Calder} A.~C.,  {Fryxell} B.,  {Plewa} T.,  {Rosner} R.,  {Dursi} L.~J.,
  {Weirs} V.~G.,  {Dupont} T.,  {Robey} H.~F.,  {Kane} J.~O.,  {Remington}
  B.~A.,  {Drake} R.~P.,  {Dimonte} G.,  {Zingale} M.,  {Timmes} F.~X.,
  {Olson} K.,  {Ricker} P.,  {MacNeice} P.,    {Tufo} H.~M.,  2002, ApJ, 143,
  201

\bibitem[\protect\citeauthoryear{{Colella} \& {Woodward}}{{Colella} \&
  {Woodward}}{1984}]{ColellaWoodward1984}
{Colella} P.,  {Woodward} P.~R.,  1984, J. Comp. Phys., 54, 174

\bibitem[\protect\citeauthoryear{{Dib}, {Brandenburg}, {Kim}, {Gopinathan} \&
  {Andr{\'e}}}{{Dib} et~al.}{2008}]{dibetal08}
{Dib} S.,  {Brandenburg} A.,  {Kim} J.,  {Gopinathan} M.,    {Andr{\'e}} P.,
  2008, ApJL, 678, L105

\bibitem[\protect\citeauthoryear{{Dimonte}, {Youngs}, {Dimits}, {Weber},
  {Marinak}, {Wunsch}, {Garasi}, {Robinson}, {Andrews}, {Ramaprabhu}, {Calder},
  {Fryxell}, {Biello}, {Dursi}, {MacNeice}, {Olson}, {Ricker} \& {and 5
  co-authors}}{{Dimonte} et~al.}{2004}]{DimonteEtAl2004}
{Dimonte} G.,  {Youngs} D.,  {Dimits} A.,  {Weber} S.,  {Marinak} M.,  {Wunsch}
  S.,  {Garasi} C.,  {Robinson} A.,  {Andrews} M.,  {Ramaprabhu} P.,  {Calder}
  A.,  {Fryxell} B.,  {Biello} J.,  {Dursi} L.,  {MacNeice} P.,  {Olson} K.,
  {Ricker} P.,    {and 5 co-authors} 2004, Physics of Fluids, 16, 1668

\bibitem[\protect\citeauthoryear{{Dubey}, {Fisher}, {Graziani}, {Jordan} IV,
  {Lamb}, {Reid}, {Rich}, {Sheeler}, {Townsley} \& {Weide}}{{Dubey}
  et~al.}{2008}]{DubeyEtAl2008}
{Dubey} A.,  {Fisher} R.,  {Graziani} C.,  {Jordan} IV G.~C.,  {Lamb} D.~Q.,
  {Reid} L.~B.,  {Rich} P.,  {Sheeler} D.,  {Townsley} D.,    {Weide} K.,
  2008, in {Pogorelov} N.~V.,  {Audit} E.,   {Zank} G.~P.,  eds, Numerical
  Modeling of Space Plasma Flows Vol.~385 of ASP Conf. Ser., p.~145

\bibitem[\protect\citeauthoryear{{Elmegreen}}{{Elmegreen}}{2008}]{elmegreen08}
{Elmegreen} B.~G.,  2008, ApJ, 672, 1006

\bibitem[\protect\citeauthoryear{{Elmegreen} \& {Falgarone}}{{Elmegreen} \&
  {Falgarone}}{1996}]{ElmegreenFalgarone1996}
{Elmegreen} B.~G.,  {Falgarone} E.,  1996, ApJ, 471, 816

\bibitem[\protect\citeauthoryear{{Elmegreen} \& {Scalo}}{{Elmegreen} \&
  {Scalo}}{2004}]{es04}
{Elmegreen} B.~G.,  {Scalo} J.,  2004, ARA\&A, 42, 211

\bibitem[\protect\citeauthoryear{{Eswaran} \& {Pope}}{{Eswaran} \&
  {Pope}}{1988}]{EswaranPope1988}
{Eswaran} V.,  {Pope} S.~B.,  1988, Computers and Fluids, 16, 257

\bibitem[\protect\citeauthoryear{{Falgarone}, {Lis}, {Phillips}, {Pouquet},
  {Porter} \& {Woodward}}{{Falgarone} et~al.}{1994}]{falgaroneetal94}
{Falgarone} E.,  {Lis} D.~C.,  {Phillips} T.~G.,  {Pouquet} A.,  {Porter}
  D.~H.,    {Woodward} P.~R.,  1994, ApJ, 436, 728

\bibitem[\protect\citeauthoryear{{Federrath}, {Klessen} \&
  {Schmidt}}{{Federrath} et~al.}{2008}]{FederrathKlessenSchmidt2008}
{Federrath} C.,  {Klessen} R.~S.,    {Schmidt} W.,  2008, ApJL, 688, L79

\bibitem[\protect\citeauthoryear{{Federrath}, {Klessen} \&
  {Schmidt}}{{Federrath} et~al.}{2009}]{FederrathKlessenSchmidt2009}
{Federrath} C.,  {Klessen} R.~S.,    {Schmidt} W.,  2009, ApJ, 692, 364

\bibitem[\protect\citeauthoryear{{Federrath}, {Roman-Duval}, {Klessen},
  {Schmidt} \& {Mac Low}}{{Federrath}
  et~al.}{2010}]{FederrathDuvalKlessenSchmidtMacLow2009}
{Federrath} C.,  {Roman-Duval} J.,  {Klessen} R.~S.,  {Schmidt} W.,    {Mac
  Low} M.,  2010, A\&A, 512, A81

\bibitem[\protect\citeauthoryear{{Fryxell}, {Olson}, {Ricker}, {Timmes},
  {Zingale}, {Lamb}, {MacNeice}, {Rosner}, {Truran} \& {Tufo}}{{Fryxell}
  et~al.}{2000}]{FryxellEtAl2000}
{Fryxell} B.,  {Olson} K.,  {Ricker} P.,  {Timmes} F.~X.,  {Zingale} M.,
  {Lamb} D.~Q.,  {MacNeice} P.,  {Rosner} R.,  {Truran} J.~W.,    {Tufo} H.,
  2000, ApJ, 131, 273

\bibitem[\protect\citeauthoryear{{Heitmann}, {Ricker}, {Warren} \&
  {Habib}}{{Heitmann} et~al.}{2005}]{HeitmannEtAl2005}
{Heitmann} K.,  {Ricker} P.~M.,  {Warren} M.~S.,    {Habib} S.,  2005, ApJ,
  160, 28

\bibitem[\protect\citeauthoryear{{Hennebelle} \& {Chabrier}}{{Hennebelle} \&
  {Chabrier}}{2008}]{HennebelleChabrier2008}
{Hennebelle} P.,  {Chabrier} G.,  2008, ApJ, 684, 395

\bibitem[\protect\citeauthoryear{{Heyer} \& {Brunt}}{{Heyer} \&
  {Brunt}}{2004}]{hb04}
{Heyer} M.~H.,  {Brunt} C.~M.,  2004, ApJL, 615, L45

\bibitem[\protect\citeauthoryear{{Hily-Blant}, {Falgarone} \&
  {Pety}}{{Hily-Blant} et~al.}{2008}]{hilyblantetal08}
{Hily-Blant} P.,  {Falgarone} E.,    {Pety} J.,  2008, A\&A, 481, 367

\bibitem[\protect\citeauthoryear{{Kitsionas}, {Federrath}, {Klessen},
  {Schmidt}, {Price}, {Dursi}, {Gritschneder}, {Walch}, {Piontek}, {Kim},
  {Jappsen}, {Ciecielag} \& {Mac Low}}{{Kitsionas}
  et~al.}{2009}]{kitsionasetal09}
{Kitsionas} S.,  {Federrath} C.,  {Klessen} R.~S.,  {Schmidt} W.,  {Price}
  D.~J.,  {Dursi} L.~J.,  {Gritschneder} M.,  {Walch} S.,  {Piontek} R.,  {Kim}
  J.,  {Jappsen} A.,  {Ciecielag} P.,    {Mac Low} M.,  2009, A\&A, 508, 541

\bibitem[\protect\citeauthoryear{{Klessen}}{{Klessen}}{2000}]{klessen00}
{Klessen} R.~S.,  2000, ApJ, 535, 869

\bibitem[\protect\citeauthoryear{{Klessen}, {Heitsch} \& {Mac Low}}{{Klessen}
  et~al.}{2000}]{khm00}
{Klessen} R.~S.,  {Heitsch} F.,    {Mac Low} M.-M.,  2000, ApJ, 535, 887

\bibitem[\protect\citeauthoryear{{Kolmogorov}}{{Kolmogorov}}{1941}]{kolmogorov%
41}
{Kolmogorov} A.,  1941, Dokl. Akad. Nauk SSSR, 30, 301

\bibitem[\protect\citeauthoryear{{Kritsuk}, {Norman}, {Padoan} \&
  {Wagner}}{{Kritsuk} et~al.}{2007}]{kritsuketal07}
{Kritsuk} A.~G.,  {Norman} M.~L.,  {Padoan} P.,    {Wagner} R.,  2007, ApJ,
  665, 416

\bibitem[\protect\citeauthoryear{{Kritsuk}, {Padoan}, {Wagner} \&
  {Norman}}{{Kritsuk} et~al.}{2007}]{kritsuketal07b}
{Kritsuk} A.~G.,  {Padoan} P.,  {Wagner} R.,    {Norman} M.~L.,  2007, in
  {Shaikh} D.,  {Zank} G.~P.,  eds, Turbulence and Nonlinear Processes in
  Astrophysical Plasmas Vol.~932 of AIP Conf. Ser., pp 393--399

\bibitem[\protect\citeauthoryear{{Krumholz} \& {McKee}}{{Krumholz} \&
  {McKee}}{2005}]{km05}
{Krumholz} M.~R.,  {McKee} C.~F.,  2005, ApJ, 630, 250

\bibitem[\protect\citeauthoryear{{Larson}}{{Larson}}{1981}]{larson81}
{Larson} R.~B.,  1981, MNRAS, 194, 809

\bibitem[\protect\citeauthoryear{{Lemaster} \& {Stone}}{{Lemaster} \&
  {Stone}}{2008}]{ls08}
{Lemaster} M.~N.,  {Stone} J.~M.,  2008, ApJL, 682, L97

\bibitem[\protect\citeauthoryear{{Mac Low} \& {Klessen}}{{Mac Low} \&
  {Klessen}}{2004}]{mk04}
{Mac Low} M.,  {Klessen} R.~S.,  2004, Rev. Mod. Phys., 76, 125

\bibitem[\protect\citeauthoryear{{Mac Low}, {Klessen}, {Burkert} \&
  {Smith}}{{Mac Low} et~al.}{1998}]{maclowetal98}
{Mac Low} M.,  {Klessen} R.~S.,  {Burkert} A.,    {Smith} M.~D.,  1998,
  Physical Review Letters, 80, 2754

\bibitem[\protect\citeauthoryear{{McKee} \& {Ostriker}}{{McKee} \&
  {Ostriker}}{2007}]{mo07}
{McKee} C.~F.,  {Ostriker} E.~C.,  2007, ARA\&A, 45, 565

\bibitem[\protect\citeauthoryear{{Monaghan}}{{Monaghan}}{1989}]{monaghan89}
{Monaghan} J.~J.,  1989, J. Comp. Phys., 82, 1

\bibitem[\protect\citeauthoryear{{Monaghan}}{{Monaghan}}{1992}]{monaghan92}
{Monaghan} J.~J.,  1992, Ann. Rev. Astron. Astrophys., 30, 543

\bibitem[\protect\citeauthoryear{{Monaghan}}{{Monaghan}}{1997}]{monaghan97}
{Monaghan} J.~J.,  1997, J. Comp. Phys., 136, 298

\bibitem[\protect\citeauthoryear{{Monaghan}}{{Monaghan}}{2005}]{monaghan05}
{Monaghan} J.~J.,  2005, Rep. Prog. Phys., 68, 1703

\bibitem[\protect\citeauthoryear{{Morris} \& {Monaghan}}{{Morris} \&
  {Monaghan}}{1997}]{mm97}
{Morris} J.~P.,  {Monaghan} J.~J.,  1997, J. Comp. Phys., 136, 41

\bibitem[\protect\citeauthoryear{{Nordlund} \& {Padoan}}{{Nordlund} \&
  {Padoan}}{1999}]{np99}
{Nordlund} {\AA}.~K.,  {Padoan} P.,  1999, in {Franco} J.,  {Carraminana} A.,
  eds, Interstellar Turbulence: Cambridge; CUP p.~218

\bibitem[\protect\citeauthoryear{{Padoan}}{{Padoan}}{1995}]{padoan95}
{Padoan} P.,  1995, MNRAS, 277, 377

\bibitem[\protect\citeauthoryear{{Padoan} \& {Nordlund}}{{Padoan} \&
  {Nordlund}}{2002}]{pn02}
{Padoan} P.,  {Nordlund} {\AA}.,  2002, ApJ, 576, 870

\bibitem[\protect\citeauthoryear{{Padoan}, {Nordlund} \& {Jones}}{{Padoan}
  et~al.}{1997}]{pnj97}
{Padoan} P.,  {Nordlund} A.,    {Jones} B.~J.~T.,  1997, MNRAS, 288, 145

\bibitem[\protect\citeauthoryear{{Padoan}, {Nordlund}, {Kritsuk}, {Norman} \&
  {Li}}{{Padoan} et~al.}{2007}]{padoanetal07}
{Padoan} P.,  {Nordlund} {\AA}.,  {Kritsuk} A.~G.,  {Norman} M.~L.,    {Li}
  P.~S.,  2007, ApJ, 661, 972

\bibitem[\protect\citeauthoryear{{Passot} \& {V{\'a}zquez-Semadeni}}{{Passot}
  \& {V{\'a}zquez-Semadeni}}{1998}]{pvs98}
{Passot} T.,  {V{\'a}zquez-Semadeni} E.,  1998, Phys. Rev. E, 58, 4501

\bibitem[\protect\citeauthoryear{{Porter}, {Woodward} \& {Pouquet}}{{Porter}
  et~al.}{1998}]{pwp98}
{Porter} D.~H.,  {Woodward} P.~R.,    {Pouquet} A.,  1998, Physics of Fluids,
  10, 237

\bibitem[\protect\citeauthoryear{{Price}}{{Price}}{2004}]{price04}
{Price} D.~J.,  2004, PhD thesis, University of Cambridge, Cambridge, UK.
  astro-ph/0507472

\bibitem[\protect\citeauthoryear{{Price}}{{Price}}{2007}]{splashpaper}
{Price} D.~J.,  2007, Publ. Astron. Soc. Aust., 24, 159

\bibitem[\protect\citeauthoryear{{Price}}{{Price}}{2010}]{price10}
{Price} D.~J.,  2010, MNRAS, 401, 1475

\bibitem[\protect\citeauthoryear{{Price} \& {Bate}}{{Price} \&
  {Bate}}{2008}]{pb08}
{Price} D.~J.,  {Bate} M.~R.,  2008, MNRAS, 385, 1820

\bibitem[\protect\citeauthoryear{{Price} \& {Monaghan}}{{Price} \&
  {Monaghan}}{2004}]{pm04b}
{Price} D.~J.,  {Monaghan} J.~J.,  2004, MNRAS, 348, 139

\bibitem[\protect\citeauthoryear{{Price} \& {Monaghan}}{{Price} \&
  {Monaghan}}{2007}]{pm07}
{Price} D.~J.,  {Monaghan} J.~J.,  2007, MNRAS, 374, 1347

\bibitem[\protect\citeauthoryear{{Salpeter}}{{Salpeter}}{1955}]{salpeter55}
{Salpeter} E.~E.,  1955, ApJ, 121, 161

\bibitem[\protect\citeauthoryear{{Schmidt}, {Federrath}, {Hupp}, {Kern} \&
  {Niemeyer}}{{Schmidt} et~al.}{2009}]{SchmidtEtAl2009}
{Schmidt} W.,  {Federrath} C.,  {Hupp} M.,  {Kern} S.,    {Niemeyer} J.~C.,
  2009, A\&A, 494, 127

\bibitem[\protect\citeauthoryear{{Schmidt}, {Federrath} \& {Klessen}}{{Schmidt}
  et~al.}{2008}]{sfk08}
{Schmidt} W.,  {Federrath} C.,    {Klessen} R.,  2008, Phys. Rev. Let., 101,
  194505

\bibitem[\protect\citeauthoryear{{Schmidt}, {Hillebrandt} \&
  {Niemeyer}}{{Schmidt} et~al.}{2006}]{SchmidtHillebrandtNiemeyer2006}
{Schmidt} W.,  {Hillebrandt} W.,    {Niemeyer} J.~C.,  2006, Computers and
  Fluids, 35, 353

\bibitem[\protect\citeauthoryear{{She} \& {Leveque}}{{She} \&
  {Leveque}}{1994}]{sl94}
{She} Z.-S.,  {Leveque} E.,  1994, Physical Review Letters, 72, 336

\bibitem[\protect\citeauthoryear{{Solomon}, {Rivolo}, {Barrett} \&
  {Yahil}}{{Solomon} et~al.}{1987}]{solomonetal87}
{Solomon} P.~M.,  {Rivolo} A.~R.,  {Barrett} J.,    {Yahil} A.,  1987, ApJ,
  319, 730

\bibitem[\protect\citeauthoryear{{Vazquez-Semadeni}}{{Vazquez-Semadeni}}{1994}%
]{vs94}
{Vazquez-Semadeni} E.,  1994, ApJ, 423, 681

\bibitem[\protect\citeauthoryear{{V{\'a}zquez-Semadeni}, {Ballesteros-Paredes}
  \& {Klessen}}{{V{\'a}zquez-Semadeni} et~al.}{2003}]{vsbpk03}
{V{\'a}zquez-Semadeni} E.,  {Ballesteros-Paredes} J.,    {Klessen} R.~S.,
  2003, ApJL, 585, L131

\bibitem[\protect\citeauthoryear{{Zuckerman} \& {Evans} II}{{Zuckerman} \&
  {Evans}}{1974}]{ze74}
{Zuckerman} B.,  {Evans} II N.~J.,  1974, ApJL, 192, L149

\end{thebibliography}

\appendix

\section{Effect of $\beta-$viscosity in the SPH calculations}
\label{sec:viscosity}
\begin{figure}
\begin{center}
\includegraphics[angle=270,width=0.95\columnwidth]{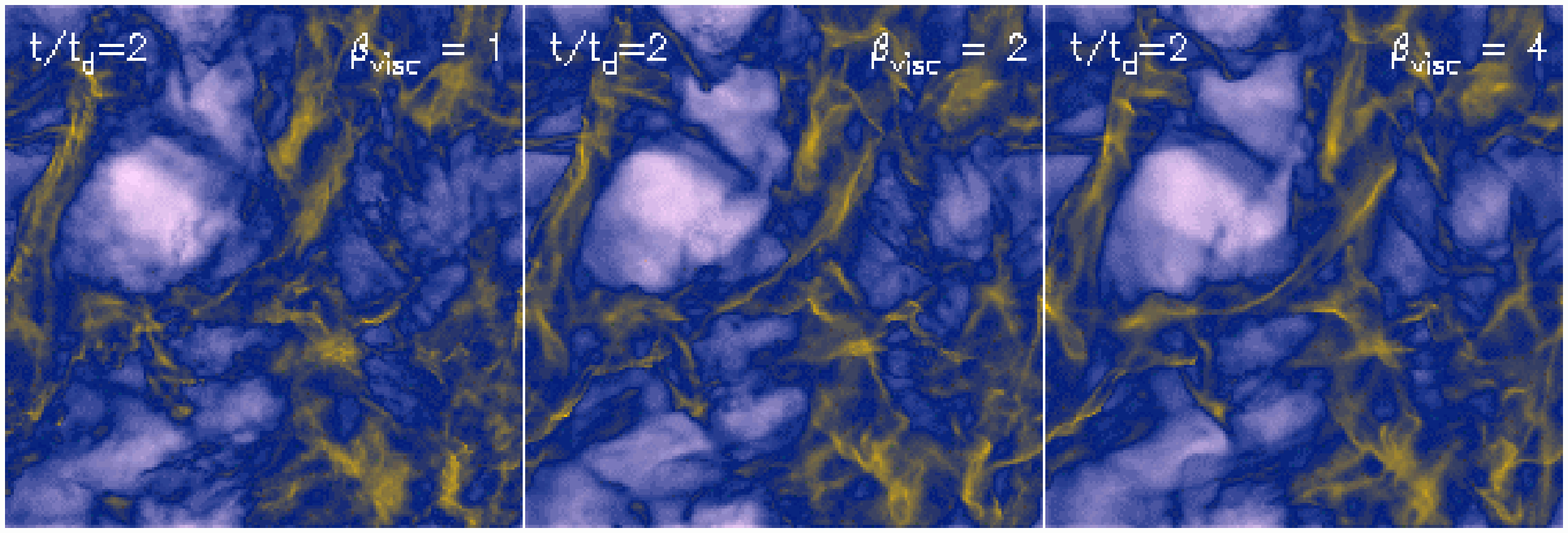}
 \caption{Effect of beta viscosity in the SPH calculations. From left to right: column density at $t/t_{d}=2$ in $128^{3}$ SPH calculations using $\beta_{visc} = 1$, $\beta_{visc}=2$ and $\beta_{visc}=4$. With $\beta_{visc}\lesssim 2$ particle penetration occurs in the shocks at these high Mach numbers, causing them to lose definition.}
\label{fig:effectofbeta}
\end{center}
\end{figure}

Fig.~\ref{fig:effectofbeta} shows the results of three $128^{3}$ particle SPH calculations after 2 dynamical times using $\beta_{visc} = 1$ (top), $\beta_{visc} = 2$ (middle) and $\beta_{visc}=4$ (bottom) in the SPH artificial viscosity term (\ref{eq:qvisc})-(\ref{eq:vsig}). At this high Mach number using $\beta_{visc} < 4$ means that particle interpenetration can occur at the shock fronts (i.e., particles overshoot the shock), resulting in excess noise in the density field compared to the grid-based results and compared to the SPH results at higher $\beta_{visc}$. In particular the shocks appear much \emph{sharper} and well-defined at higher $\beta_{visc}$, somewhat counter intuitively since we are adding more viscosity. Using $\beta_{visc}=4$ shows good agreement with the grid-based shock structures (c.f. Figs.~\ref{fig:coldens512} and \ref{fig:slicerho_t1}).

\section{Effect of fitting range on the measured structure function slope}
\label{sec:sfslopefit}
\begin{figure}
\begin{center}
\includegraphics[width=\columnwidth]{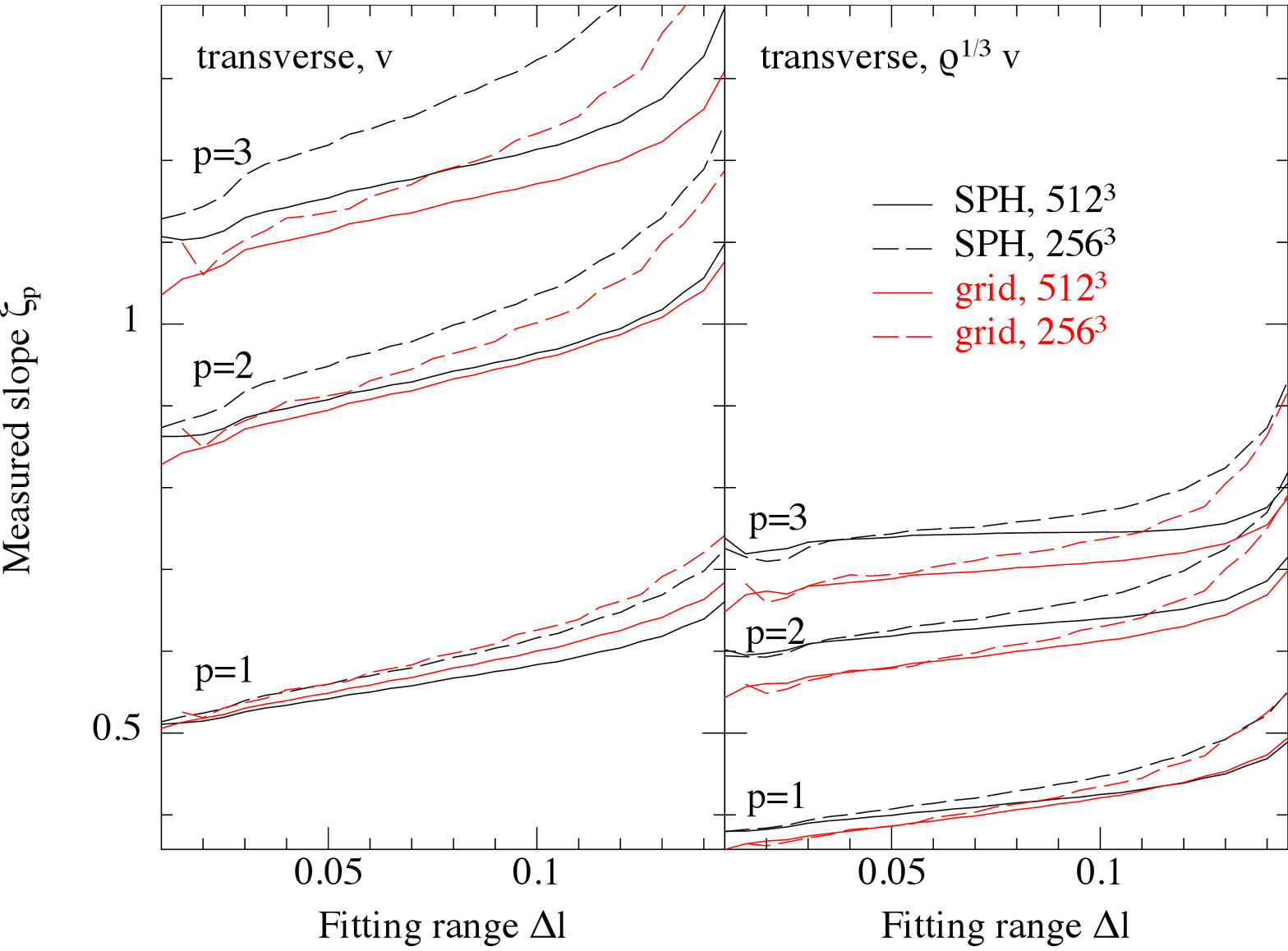}
 \caption{The dependence of the measured structure function slope $\zeta_{p}$ on the range in scales (lag) over which the least squares fit to the time averaged structure function is performed, where $\Delta l \equiv l_{max} - l_{min}$ and we have used a fixed $l_{max} = 0.15$. The measured slope, shown for the three lowest-order structure functions ($p=1,2,3$) shows a strong dependence on the adopted fitting range, particularly for the velocity structure functions (left panel) and to a lesser extent for the density-weighted $(\rho^{1/3} {\bf v}$) versions (right panel). Thus, although the slopes appear to converge in individual codes for a fixed $\Delta l$, there remains a strong dependence on $\Delta l$, indicating that at a scaling range of constant $\zeta_{p}$ is not resolved in either code at the resolutions employed.}
\label{fig:slopefitting}
\end{center}
\end{figure}
The best-fit structure function slopes, $\zeta_{p}$, computed from a least-squares fit to the transverse velocity and density-weighted structure functions shown in Fig.~\ref{fig:sfscalev}, are shown in Fig.~\ref{fig:slopefitting} as a function of the range in the lag $\Delta l \equiv l_{max} - l_{min}$ used to perform the fitting. To produce the Figure we initially adopted a fixed $l_{max} = 0.15$, based on visual inspection of Fig.~\ref{fig:sfall} for the transverse case, and performed the fit down to $l_{min} = l_{max} - \Delta l$, plotting the resulting slope as a function of $\Delta l$. The results show a strong steepening of the measured slope as the fitting range is increased, indicating that the slope is not well-fitted by a single power-law. The $\rho^{1/3}$-weighted structure functions (right panel) show a weaker dependence on $\Delta l$ than for the pure velocity equivalents (left panel), though the results from the two codes even in this case do not show convergence with each other. Convergence in each individual code is only obtained if a relatively short fitting range $\Delta l \lesssim 0.05$ is adopted, but it is clear that the converged value will nevertheless change if a different $\Delta l$ is used. Neither can it be asserted that the adopted $\Delta l$ should be changed with resolution, since it is not clear that any scaling range has been resolved even at the highest resolution adopted in either code (making the ``correct'' choice $\Delta l = 0$ for the numerical resolutions tested in this code comparison).

\label{lastpage}
\enddocument